\documentclass[prc,letterpaper,twocolumn,showpacs,showkeys,lengthcheck,tightenlines,
               nofootinbib,preprintnumbers,superscriptaddress]{revtex4-1} 

\pdfpageattr {/Group << /S /Transparency /I true /CS /DeviceRGB>>}

\usepackage[utf8]{inputenc}
\usepackage{newtxtext,newtxmath}

\usepackage{titlesec}
\usepackage{dcolumn}
\usepackage{bm}
\usepackage{amsmath,amssymb,amsfonts}

\usepackage{graphicx}

\usepackage{color}
\usepackage{bbold}

\usepackage{slashed}
\usepackage[svgnames]{xcolor}
\usepackage[english]{babel}
\usepackage{blindtext}
\usepackage{microtype}
\usepackage{tikz}

\usepackage[colorlinks=true,linkcolor=blue,urlcolor=blue,citecolor=blue]{hyperref}
\usepackage{ifpdf}

\def\a{\alpha}
\def\b{\beta}
\def\d{\delta}
\def\g{\gamma}

\def\ve{\varepsilon}

\def\l{\lambda}

\def\G{\Gamma}
\def\L{\Lambda}

\def\pl{\partial}
\def\hs{\hspace}

\def\no{\nonumber}

\def\lf{\left}
\def\rg{\right}

\newcommand{\ph}[1]{\phantom{#1}}

\newcommand{\sh}[1]{\slashed{#1}}

\font\bb=bbmss10 scaled 1200
\def\ident{\mbox{\bb 1}}

\graphicspath{{.}{figures/}} 

\titlespacing{\section}{4pt}{10pt plus 4pt minus 2pt}{8pt plus 2pt minus 2pt}
\titlespacing{\subsection}{0pt}{12pt plus 4pt minus 2pt}{8pt plus 2pt minus 2pt}

\usepackage[mathscr,scaled=1.15]{urwchancal}
\DeclareFontFamily{OT1}{pzc}{}
\DeclareFontShape{OT1}{pzc}{m}{it}%
{<-> s * [1.15] pzcmi7t}{}
\DeclareMathAlphabet{\mathpzc}{OT1}{pzc}{m}{it}

\makeatletter
\newcommand*\if@single[3]{%
  \setbox0\hbox{${\mathaccent"0362{#1}}^H$}%
  \setbox2\hbox{${\mathaccent"0362{\kern0pt#1}}^H$}%
  \ifdim\ht0=\ht2 #3\else #2\fi
  }
\newcommand*\rel@kern[1]{\kern#1\dimexpr\macc@kerna}
\newcommand*\widebar[1]{\@ifnextchar^{{\wide@bar{#1}{0}}}{\wide@bar{#1}{1}}}
\newcommand*\wide@bar[2]{\if@single{#1}{\wide@bar@{#1}{#2}{1}}{\wide@bar@{#1}{#2}{2}}}
\newcommand*\wide@bar@[3]{%
  \begingroup
  \def\mathaccent##1##2{%
    \if#32 \let\macc@nucleus\first@char \fi
    \setbox\z@\hbox{$\macc@style{\macc@nucleus}_{}$}%
    \setbox\tw@\hbox{$\macc@style{\macc@nucleus}{}_{}$}%
    \dimen@\wd\tw@
    \advance\dimen@-\wd\z@
    \divide\dimen@ 3
    \@tempdima\wd\tw@
    \advance\@tempdima-\scriptspace
    \divide\@tempdima 10
    \advance\dimen@-\@tempdima
    \ifdim\dimen@>\z@ \dimen@0pt\fi
    \rel@kern{0.6}\kern-\dimen@
    \if#31
      \overline{\rel@kern{-0.6}\kern\dimen@\macc@nucleus\rel@kern{0.4}\kern\dimen@}%
      \advance\dimen@0.4\dimexpr\macc@kerna
      \let\final@kern#2%
      \ifdim\dimen@<\z@ \let\final@kern1\fi
      \if\final@kern1 \kern-\dimen@\fi
    \else
      \overline{\rel@kern{-0.6}\kern\dimen@#1}%
    \fi
  }%
  \macc@depth\@ne
  \let\math@bgroup\@empty \let\math@egroup\macc@set@skewchar
  \mathsurround\z@ \frozen@everymath{\mathgroup\macc@group\relax}%
  \macc@set@skewchar\relax
  \let\mathaccentV\macc@nested@a
  \if#31
    \macc@nested@a\relax111{#1}%
  \else
    \def\gobble@till@marker##1\endmarker{}%
    \futurelet\first@char\gobble@till@marker#1\endmarker
    \ifcat\noexpand\first@char A\else
      \def\first@char{}%
    \fi
    \macc@nested@a\relax111{\first@char}%
  \fi
  \endgroup
}
\makeatother

\begin{document}
\title{Charge Symmetry Breaking Effects in Pion and Kaon Structure}

\author{Parada~T.~P.~Hutauruk}
\affiliation{Asia Pacific Center for Theoretical Physics, Pohang, Gyeongbuk 37673, South Korea}

\author{Wolfgang Bentz}
\affiliation{Department of Physics, School of Science, Tokai University,
             4-1-1 Kitakaname, Hiratsuka-shi, Kanagawa 259-1292, Japan}
\affiliation{Radiation Laboratory, Nishina Center, RIKEN, Wako, Saitama 351-0198,
Japan}

\author{Ian~C.~Clo{\"e}t}
\affiliation{Physics Division, Argonne National Laboratory, Argonne, Illinois 60439, USA}

\author{Anthony~W.~Thomas}
\affiliation{CSSM and ARC Centre of Excellence for Particle Physics at the Terascale, \\ 
             Department of Physics, University of Adelaide, Adelaide SA 5005, Australia}

\preprint{ADP-18-2/T1050}

\begin{abstract}
Charge symmetry breaking (CSB) effects associated with the $u$ and $d$ quark mass difference are investigated in the quark distribution functions and spacelike electromagnetic form factors of the pion and kaon. We use a confining version of the Nambu--Jona-Lasinio model, where CSB effects at the infrared scale associated with the model are driven by the dressed $u$ and $d$ quark mass ratio, which because of dynamical chiral symmetry breaking is much closer to unity than the associated current quark mass ratio. The pion and kaon are given as bound states of a dressed quark and a dressed antiquark governed by the Bethe-Salpeter equation, and exhibit the properties of Goldstone bosons, with a pion mass difference given by $m_{\pi^+}^2 - m_{\pi^0}^2 \propto (m_u - m_d)^2$ as demanded by dynamical chiral symmetry breaking. We find significant CSB effects for realistic current quark mass ratios ($m_u/m_d \sim 0.5$) in the quark flavor-sector electromagnetic form factors of both the pion and kaon. For example, the difference between the $u$ and $d$ quark contributions to the $\pi^+$ electromagnetic form factors is about 8\% at a momentum transfer of $Q^2 \simeq 10\,$GeV$^2$, while the analogous effect for the light quark sector form factors in the $K^+$ and $K^0$ is about twice as large. For the parton distribution functions we find CSB effects which are considerably smaller than those found in the electromagnetic form factors. 
\end{abstract}


\maketitle
\section{INTRODUCTION}
Charge symmetry breaking (CSB) provides a powerful tool with which to study and understand strong interaction systems~\cite{Miller:1990iz,Slaus:1990nn,Miller:2006tv,Londergan:2009kj}. In quantum chromodynamics (QCD) CSB effects result from the mass difference between the $u$ and $d$ quarks, while the difference in the $u$ and $d$ quark electric charges is the dominant electroweak effect~\cite{Londergan:2009kj}. Empirically, CSB effects are clearly evident in the proton-neutron mass difference, and the differing masses between the charged and neutral pion and kaon states, where for the pion the difference is purely electromagnetic up to $\mathcal{O}[(m_u - m_d)^2]$ corrections~\cite{Gasser:1982ap}. CSB effects in hadron masses have been studied using dynamical lattice simulation of QED+QCD~\cite{Borsanyi:2013lga,Horsley:2015vla}, where, for example, the QCDSF-UKQCD collaboration found {\it inter alia} that the QCD CSB effects between the kaons are much larger than between the proton and neutron~\cite{Horsley:2015eaa}.

In a different area, CSB is an important background in the extraction of the strange electromagnetic form factor and parton distribution functions (PDFs) of the nucleon~\cite{Kubis:2009cp}, where for the former a combination of lattice QCD and effective field theory has led to a significant increase in the precision with which that background is known~\cite{Shanahan:2015caa}. As a final example, we note that CSB in the PDFs of the nucleon is vital to understanding the NuTeV anomaly~\cite{Zeller:2001hh,Cloet:2009qs} and this has led to a number of studies~\cite{Martin:2003sk,Londergan:2005ht,Bentz:2009yy,Wang:2015msk}. In particular, lattice results~\cite{Shanahan:2013vla}  are in agreement with much earlier calculations within the MIT bag model~\cite{Rodionov:1994cg,Sather:1991je,Londergan:1998ai,Londergan:2009kj} for CSB associated with the quark mass difference. Recent work has also brought the QED contribution to CSB in PDFs under better control~\cite{Wang:2015msk}.

Beyond mass differences and effects in low energy nuclear physics~\cite{Miller:2006tv}, such as the Nolen-Schiffer anomaly~\cite{Nolen:1969ms,Henley:1989vi,Hatsuda:1990pj,Saito:1994tq}, the experimental study of CSB effects is challenging. Definitive experiments are certainly needed, where promising examples include parity-violating deep inelastic scattering (DIS) on the deuteron~\cite{Miller:2013nea} and $\pi^+/\pi^-$ production in semi-inclusive DIS from the nucleon~\cite{Londergan:1996vf}, both of which are planned at Jefferson Lab. In addition, interesting possibilities exist at an electron-ion collider~\cite{Accardi:2012qut}, such as charged current reactions~\cite{Boer:2011fh}, and using pion-induced Drell-Yan reactions~\cite{Londergan:1994gr}. In this work we investigate the effect of CSB arising from the $u$ and $d$ quark mass difference in the leading-twist PDFs and electromagnetic form factors of the pion and kaon. This study is performed in the expectation that the size of CSB effects can be better understood and estimated, and, for example, because knowledge of these effects is essential to accurately extract the CSB effects, and $s$-quark content, in the nucleon through processes such as pion-induced Drell-Yan~\cite{Conway:1989fs,Londergan:1994gr}. To perform these calculations we use the Nambu--Jona-Lasinio (NJL) model~\cite{Nambu:1961fr,Nambu:1961tp,Klevansky:1992qe,Vogl:1991qt,Cloet:2014rja}, regularized using the proper-time scheme~\cite{Schwinger:1951nm} so that quark confinement effects are included~\cite{Ebert:1996vx,Hellstern:1997nv,Bentz:2001vc}. The outline of the paper is as follows: Sect.~\ref{sec:njl} presents the theoretical framework used to study CSB, and results are given for CSB effects on masses and effective couplings; Sect.~\ref{sec:csb} presents results for CSB effects in the pion and kaon spacelike electromagnetic form factors and PDFs; and Sect.~\ref{sec:summary} provides a summary.

\section{CSB AND THE NJL MODEL\label{sec:njl}}
The NJL model is quark-level chiral effective field theory of QCD and has been used with success to describe numerous non-perturbative phenomena in QCD~\cite{Vogl:1989ea,Klimt:1989pm,Ishii:1995bu,Bentz:2001vc,Cloet:2007em,Carrillo-Serrano:2014zta,Ito:2009zc,Matevosyan:2011vj,Cloet:2012td,Hutauruk:2016sug,Ninomiya:2017ggn,Bernard:1989fe,Bernard:1988bx,Lemmer:1995eb,Schulze:1994fy,Weiss:1993kv,Blin:1987hw,Theussl:2002xp,Noguera:2015iia,Courtoy:2007vy,Hippe:1995hu,Shigetani:1993dx,Davidson:1994uv,Davidson:2001cc,RuizArriola:2002wr,Dmitrasinovic:1992hb}. Its key features are that it shares the same global symmetries as QCD, and is a Poincar\'e covariant quantum field theory that exhibits dynamical chiral symmetry breaking. The NJL model therefore naturally describes the appearance of a non-zero quark condensate, which is directly linked to dynamically generated dressed quark masses of several hundred MeV (even in the chiral limit), and pions and kaons as Goldstone bosons. The NJL model is therefore an ideal tool with which to study CSB effects in hadron structure because it is the dressed quark masses, not the current quark masses, that determine the size of CSB at scales similar to $\L_{\text{QCD}}$.

The three-flavor NJL Lagrangian, containing only four-fermion interaction terms, has the form\footnote{To express the Lagrangian in this compact form we have chosen the couplings of the flavor-singlet pieces of the $G_\rho$ term to equal $2/3$ times the flavor-octet coupling (note that $\lambda_0 \equiv \sqrt{2/3}\,\scalebox{0.8}{$\ident$}$). Such a choice avoids flavor mixing, giving the flavor content of the $\omega$ meson as $u \bar{u} + d \bar{d}$ and the $\phi$ meson as $s\bar{s}$, and the $\omega$ and $\rho$ are mass degenerate.}
\begin{align}
\mathcal{L}_{NJL} &= \bar{\psi}(i\slashed{\partial} 
- \hat{m})\psi  
+ G_{\pi}\left[(\bar{\psi}\,\lambda_{a}\,\psi)^{2} 
- (\bar{\psi}\,\lambda_{a}\,\gamma_{5}\,\psi)^{2}\,\right] \no \\
&\hs{10mm}
- G_\rho\left[(\bar{\psi}\,\lambda_{a}\,\gamma^\mu\,\psi)^{2} 
+ (\bar{\psi}\,\lambda_{a}\,\gamma^\mu \gamma_{5}\,\psi)^{2}\right],
\label{eq:njl_lagrangian}
\end{align}
where the quark field has the flavor components $\psi = (u, d, s)$,  $\hat{m} = {\rm diag}(m_u, m_d, m_s)$ denotes the current quark mass matrix, $G_{\pi}$ and $G_\rho$ are four-fermion  coupling constants, and $\lambda_0,\dots,\lambda_8$ are the Gell-Mann  matrices in flavor space where $\lambda_0 \equiv \sqrt{2/3}\,\ident$.  The NJL model has divergences, which we choose to regularize using the proper-time regularization scheme with an infrared cutoff, because it simulates aspects of quark confinement by eliminating on-shell quark propagation~\cite{Cloet:2014rja,Hutauruk:2016sug}.

The dressed quark mass $M_q$ for each quark flavor $q = u, d, s$, is determined by evaluating the gap equation, which in the proper-time regularization scheme takes the form~\cite{Cloet:2014rja,Hutauruk:2016sug}
\begin{align}
M_q &= m_q + \frac{3\,G_\pi\,M_q}{\pi^2} \int_{1/\Lambda_{\rm UV}^2}^{1/\Lambda_{\rm IR}^2} 
\frac{d\tau}{\tau^2}\ e^{-\tau\,M_q^2},
\label{eq:gap}
\end{align}
where $m_q$ is the current quark mass of flavor $q$, and $\Lambda_{\rm IR},\,\Lambda_{\rm UV}$ are respectively the infrared and ultraviolet proper-time regularization parameters. It is clear that $\Lambda_{\rm UV}$ removes the poles at $\tau=0$ and renders the theory finite, while $\Lambda_{\rm IR}$ removes particle propagation for large values of the proper-time parameter $\tau$~\cite{Schwinger:1951nm}, thereby simulating aspects of quark confinement~\cite{Ebert:1996vx,Hellstern:1997nv}. Equation~\eqref{eq:gap} demonstrates that with only four-fermion interactions in the Lagrangian there is no flavor mixing in the gap equation. In forthcoming results we will drop the regularization parameters to aid clarity.

The pion and kaon are given as relativistic bound-states of a dressed-quark and a dressed-antiquark whose properties are determined by solving the $\bar{q}q$ Bethe-Salpeter equation (BSE) in the pseudoscalar channel~\cite{Cloet:2014rja,Hutauruk:2016sug}. With the Lagrangian of Eq.~\eqref{eq:njl_lagrangian} the $t$-matrix solution to the BSE has the form
\begin{align}
\l_\a\g_5\ \tau_\a (q)\ \l_\a^\dagger\g_5 &= \l_\a\g_5\ \frac{-2i\,G_\pi}{1 + 2\,G_\pi\,\Pi_\a (q^2)}\ \l_\a^\dagger\g_5,
\label{eq:tmatrix}
\end{align}
where $\a=\pi^{\pm},\,\pi^0,\,K^{\pm},\,K^0,\bar{K}^0$ and the bubble diagrams take the form:
\begin{align}
\label{eq:bubblegraphtot}
\Pi_{\a}(q^2) &= i\! \int\! \frac{d^4k}{(2\pi)^4}\ 
\mathrm{Tr}\!\left[\g_5\,\l_\a^\dagger\,S(k)\,\g_5\,\l_\a\,S(k+q) \right],
\end{align}
where the trace is over Dirac, color and flavor indices.
The flavor matrices for the relevant meson channels are $\l_{\pi^\pm} = \frac{1}{\sqrt{2}}\lf(\l_1 \pm i\l_2\rg)$, $\l_{\pi^0} = \l_3$, $\l_{K^\pm} = \frac{1}{\sqrt{2}}\lf(\l_4 \pm i\l_5\rg)$, $\l_{K^0} = \frac{1}{\sqrt{2}}\lf(\l_6 + i\l_7\rg)$, $\l_{\bar{K}^0} = \frac{1}{\sqrt{2}}\lf(\l_6 - i\l_7\rg)$,\footnote{The flavor matrices are normalized such that Tr$\,[\l^\dagger_\a\l_\b]=2\,\delta_{\a\b}$.} and the dressed quark propagator is diagonal in flavor space: $S(k) = \text{diag}\lf[S_u(k),\,S_d(k),\,S_s(k)\rg]$, where $S_q^{-1} (k) = \sh{k} - M_q + i \ve$. The masses of the pseudoscalar mesons are given by the poles in the appropriate $t$-matrix of Eq.~\eqref{eq:tmatrix}, that is
\begin{align}
1 + 2\,G_\pi\,\Pi_\a (q^2 = m_\a^2) = 0.
\end{align}
By using the gap equation the following relation can be obtained:
\begin{align}
m_\a^2 &= \left[ \frac{m_i}{M_i} + \frac{m_j}{M_j} \right] 
\frac{1}{G_\pi\, \mathcal{I}_{ij}(m_\a^2)} + (M_i - M_j)^2,
\end{align}
which holds for all $\a$, with the exception of $\a = \pi^0$ which is given by
\begin{align}
m_{\pi^0}^2 &= \frac{m_u}{M_u}\,\frac{1}{G_\pi\, \mathcal{I}_{uu}(m_{\pi^0}^2)} + \frac{m_d}{M_d}\,\frac{1}{G_\pi\, \mathcal{I}_{dd}(m_{\pi^0}^2)},
\end{align}
where in both cases
\begin{align}
\mathcal{I}_{ij}(k^2) &= \frac{3}{\pi^2} \int_0^1\! dx\! \int
\! \frac{d\tau}{\tau}\,
e^{-\tau\lf[x(x-1)\,k^2 + x\,M_j^2 + (1-x)\,M_i^2\rg]}.
\end{align} 
Note, $M_{i,j}$ are the dressed quark masses that appear in the meson $\a$.
These results illustrate the Goldstone boson nature of the pion and kaon. In addition, by using the gap equation it is straightforward to show that the mass splitting between the neutral and charged pions caused by $m_u \neq m_d$ is quadratic in the current-quark mass difference, that is, $m_{\pi^\pm}^2 - m_{\pi^0}^2 \propto (m_u - m_d)^2$, as required by dynamical chiral symmetry breaking~\cite{Gasser:1982ap}. Near a bound state pole the $t$-matrix behaves as
\begin{align}
\l_\a\g_5\ \tau_\a (q)\ \l_\a^\dagger\g_5 &\to \frac{Z_\a\, \l_\a\g_5\  \l_\a^\dagger\g_5}{p^2 - m_\a^2 + i\ve},
\end{align}
which defines the homogeneous Bethe-Salpeter vertices~\cite{Cloet:2014rja,Hutauruk:2016sug}:
\begin{align}
\G_\a &= \sqrt{Z_\a}\,\g_5\,\l_\a, \qquad \widebar{\G}_\a = \sqrt{Z_\a}\,\g_5\,\l_\a^\dagger,
\end{align}
where the residue at the pole is given by
\begin{align}
\label{eq:couplinconstant}
Z_{\a}^{-1} &= -\left.\frac{\partial\, \Pi_\a(q^2)}{\partial q^2} \right|_{q^2 = m_\a^2}.
\end{align}
It is standard to interpret $\sqrt{Z_\a}$ as the effective meson-quark-quark coupling constant.

\begin{table*}[tbp]
\addtolength{\tabcolsep}{2.5pt}
\addtolength{\extrarowheight}{2.2pt}
\begin{tabular}{cccccccccccccccccccc}
\hline\hline
$m_u/m_d$ & $m_u$ & $m_d$ & $M_u$ & $M_d$ & $m_{\pi^0}$ & $m_{K^\pm}$ & $m_{K^0}$ & $m_{\rho^+}$ & $f_{\pi^0}$ & $Z_{\pi^0}$ & $Z_{\pi^\pm}$ & $Z_{K^\pm}$ & $Z_{K^0}$ & $G_\pi$ & $G_\rho$ & $\L_{\rm UV}$ \\[0.2em]
\hline
0        & 0    & 32.9 & 387 & 412 & 137.84 & 483 & 507 & 775.67 & 92.83 & 17.830 & 17.842 & 20.73 & 21.04 & 19.06 & 10.731 & 644.52 \\
0.1      & 2.99 & 29.9 & 390 & 410 & 138.56 & 486 & 504 & 775.44 & 92.89 & 17.837 & 17.846 & 20.76 & 21.01 & 19.05 & 10.746 & 644.64 \\
0.3      & 7.58 & 25.3 & 393 & 406 & 139.38 & 489 & 501 & 775.19 & 92.95 & 17.846 & 17.850 & 20.80 & 20.97 & 19.05 & 10.764 & 644.77 \\
0.5      & 11.0 & 21.9 & 396 & 404 & 139.76 & 491 & 499 & 775.07 & 92.98 & 17.850 & 17.852 & 20.83 & 20.94 & 19.05 & 10.773 & 644.83 \\
0.7      & 13.5 & 19.3 & 398 & 402 & 139.93 & 493 & 497 & 775.02 & 92.99 & 17.852 & 17.853 & 20.86 & 20.91 & 19.04 & 10.776 & 644.86 \\
0.9      & 15.6 & 17.3 & 399 & 401 & 139.99 & 494 & 496 & 775.00 & 93.00 & 17.853 & 17.853 & 20.88 & 20.89 & 19.04 & 10.778 & 644.87 \\
1        & 16.4 & 16.4 & 400 & 400 & 140    & 495 & 495 & 775    & 93    & 17.853 & 17.853 & 20.89 & 20.89 & 19.04 & 10.778 & 644.87 \\
\hline\hline
\end{tabular}
\caption{Results for the current and dressed quark masses, neutral pion, kaon and $\rho^+$ masses, neutral pion leptonic decay constant, meson-quark-quark coupling constants, and the model parameters that vary with $m_u/m_d$. Recall, that the mass and decay constant of the charged pions, and the $\rho^0$ mass, are fixed at their physical values and therefore do not vary with $m_u/m_d$. Similarly, the strange quark mass is keep constant as CSB effects are introduced. Note, dimensioned quantities are in units of MeV, with the exception of $G_{\pi,\rho}$ which are in units of GeV$^{-2}$.}
\label{tab:parameters}
\end{table*}

\begin{figure}[bp]
\centering\includegraphics[width=\columnwidth]{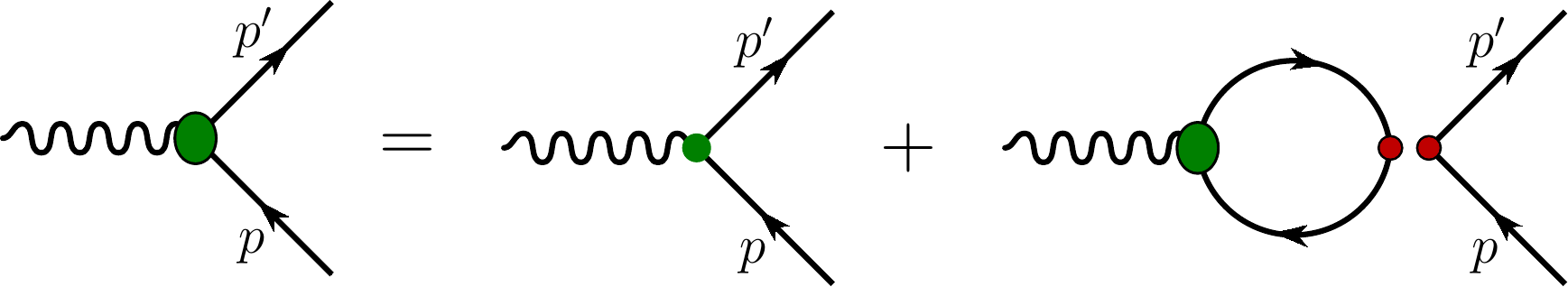}
\caption{(Colour online) The inhomogeneous BSE which gives the dressed quark-photon vertex.  
The large shaded oval represents the solution of the inhomogeneous BSE, the small dot is the inhomogeneous driving term ($\hat{Q}\,\g^\mu$) and the double-dots represent the $q\bar{q}$ interaction kernel derived from the NJL Lagrangian~\cite{Cloet:2014rja}.}
\label{fig:vectormesons}
\end{figure}

We are interested in CSB effects in electromagnetic form factors, and as such an essential ingredient is the dressed quark-photon vertex, which is given by the solution to the inhomogeneous BSE illustrated in Fig.~\ref{fig:vectormesons}. With the NJL Lagrangian of Eq.~\eqref{eq:njl_lagrangian} and the associated quark--antiquark interaction kernel (see Eq.~(2) of Ref.~\cite{Hutauruk:2016sug}), the general solution for the dressed quark-photon vertex in flavor space has the form
\begin{align}
\label{eq:quarkvertex}
\L^\mu_{\g\,Q}(p',p) &= \hat{Q}\,\g^\mu + \lf(\g^\mu - \frac{q^\mu\sh{q}}{q^2}\rg)F_{Q}(Q^2),
\end{align}
where $q^2 = (p'-p)^2 \equiv -Q^2$, $\hat{Q} = \text{diag}\lf[e_u,\,e_d,\,e_s\rg]$ is the quark charge operator and $F_{Q}(Q^2) = \text{diag}\lf[e_u\,F_U(Q^2),\,e_d\,F_D(Q^2),\,e_s\,F_S(Q^2)\rg]$ contains the dressed quark form factors. This result clearly satisfies the Ward-Takahashi identity:
\begin{align}
q_\mu\,\L^\mu_{\g\,Q}(p',p) = \hat{Q}\lf[S^{-1}(p') - S^{-1}(p)\rg],
\end{align}
and therefore respects electromagnetic gauge invariance. For the dressed quark form factors we find
\begin{align}
\label{eq:f1U}
F_{U,D,S}(Q^2) &= \frac{-2\,G_\rho\,\Pi_{v}^{u,d,s}(Q^2)}{1 + 2\,G_\rho\,\Pi_{v}^{u,d,s}(Q^2)},
\end{align}
where the explicit form of the bubble diagram is
\begin{align}
\hs*{-1mm}\Pi^{q}_{v} (Q^2) &= \frac{3\,Q^2}{\pi^2}\! \int_0^1\!\! dx \int\!\! \frac{d\tau}{\tau}\
x\lf(1-x\rg)\, e^{-\tau\lf[M_q^2 + x\lf(1-x\rg)Q^2\rg]}.
\end{align}
For later convenience we define the form factors
\begin{align}
F_{1Q}(Q^2) = e_q\,\lf[1 + F_Q(Q^2)\rg] = e_q\,f_{1Q}(Q^2),
\label{eq:chargeffs}
\end{align}
which may be interpreted as the dressed quark charge form factors, with $Q=U,D,S$. Therefore, with the NJL Lagrangian of Eq.~\eqref{eq:njl_lagrangian} there is no flavor mixing in the dressed quark form factors, in analogy with the dressed quark masses. In the limit $Q^2 \to \infty$ these form factors reduce to the elementary quark charges, as expected because of asymptotic freedom in QCD, and for small $Q^2$ these results are similar to expectations from vector meson dominance, where the dressed $u$ and $d$ quarks are dressed by $\rho$ and $\omega$ mesons, and the $s$ quark by the $\phi$ meson~\cite{Cloet:2014rja}. Note, the denominators in Eq.~\eqref{eq:f1U} are the same as the pole condition obtained by solving the BSE in the $\rho$, $\omega$ or $\phi$ channels, and therefore these form factors have poles in the timelike region at the associated meson mass.

The model therefore has the following parameters: the current quark masses $m_u$, $m_d$ and $m_s$; the four-fermion coupling constants $G_\pi$ and $G_\rho$; and the regularization parameters $\L_{\rm IR}$ and $\L_{\rm UV}$. To expose the effects of CSB as clearly as possible we choose to fit these parameters as follows: for a given current-quark mass ratio $m_u/m_d$ we constrain $G_\pi$, $G_\rho$ and $\L_{\rm UV}$ by the physical values for the $\pi^+$ mass and leptonic decay constant, and the $\rho^0$ mass, which we take to equal $m_{\pi^+} = 140\,$MeV, $m_{\rho^0} = 775\,$MeV and $f_{\pi^+} = 93\,$MeV.\footnote{For explicit expressions that give these observables see Ref.~\cite{Cloet:2014rja}.} The remaining parameters are fit in the charge symmetric limit ($m_u = m_d$) and do not change when CSB effects are introduced. The strange current quark mass is constrained to give the physical kaon mass ($m_K = 495\,$MeV), the infrared cutoff sets the confinement scale of the model and therefore we set $\Lambda_{\rm IR} = 240\,$MeV$\,\simeq \L_{\rm QCD}$, and finally the average current quark mass $m_0 = \frac{1}{2}\lf(m_u + m_d\rg)$ is constrained to give a dressed-mass of $M_0 = 400\,$MeV in the charge symmetric limit, in agreement with previous studies~\cite{Cloet:2014rja,Hutauruk:2016sug}. 

The current quark mass ratio, $r_{ud} = m_u/m_d$, is a free parameter, which we adjust to study CSB effects. For a given $r_{ud}$ we have
\begin{align}
m_{u,d} &= m_0 \mp \d m, \qquad \text{where} \qquad \d m = m_0\,\frac{1 - r_{ud}}{1+r_{ud}}.
\end{align}
With these constraints we find $m_s = 356\,$MeV ($M_s = 611\,$MeV) and $m_0 = 16.4\,$MeV and therefore $m_s/m_0 = 21.7$ which is in reasonable agreement with the empirical value of $2\,m_s / (m_u + m_d) = 27.5\pm 1.0$~\cite{Beringer:1900zz,Durr:2010vn}, and for the $\phi$ meson mass we have $m_\phi = 1001\,$MeV which is within 2\% of the physical value ($m_\phi \simeq 1019\,$MeV). Further results for quark and meson masses, model parameters, and coupling constants for various values of $m_u/m_d$ are given in Tab.~\ref{tab:parameters}.

For a realistic current quark mass ratio of $m_u/m_d = 0.5$~\cite{Patrignani:2016xqp} we have CSB effects in the current quarks of $(m_d-m_u)/(m_u+m_d) = 33\%$, whereas for the dressed quark masses we have $(M_d-M_u)/(M_u+M_d) = 1\%$. It is therefore clear that in the infrared dynamical chiral symmetry breaking (DCSB) dramatically reduces the size of CSB effects that may be expected from the current quark masses. For the same $m_u/m_d$ ratio we find $m_{\pi^\pm} - m_{\pi^0} = 0.24\,$MeV which is much smaller than the empirical value of $4.59\,$MeV~\cite{Patrignani:2016xqp}, and is therefore in agreement with expectation that this mass-splitting is dominated by QED effects. For the kaon mass we have $(m_{K^0} - m_{K^\pm})/(m_{K^0} + m_{K^\pm}) = 0.8\%$ with $m_{K^0} - m_{K^\pm} = 7.8\,$MeV which is about twice the empirical splitting of $3.93\,$MeV, and therefore QED effects must reduce this mass splitting, which is the finding of lattice calculations~\cite{Horsley:2015eaa}. Finally, the CSB effects in the effective meson-quark-quark coupling and the pion's leptonic decay constant are shown to be negligible (see Table.~\ref{tab:parameters}), and the same is found for the chiral condensates.

\section{CSB IN PSEUDOSCALAR MESON FORM FACTORS AND PDFS \label{sec:csb}}
\begin{figure}[tbp]
\centering\includegraphics[width=\columnwidth]{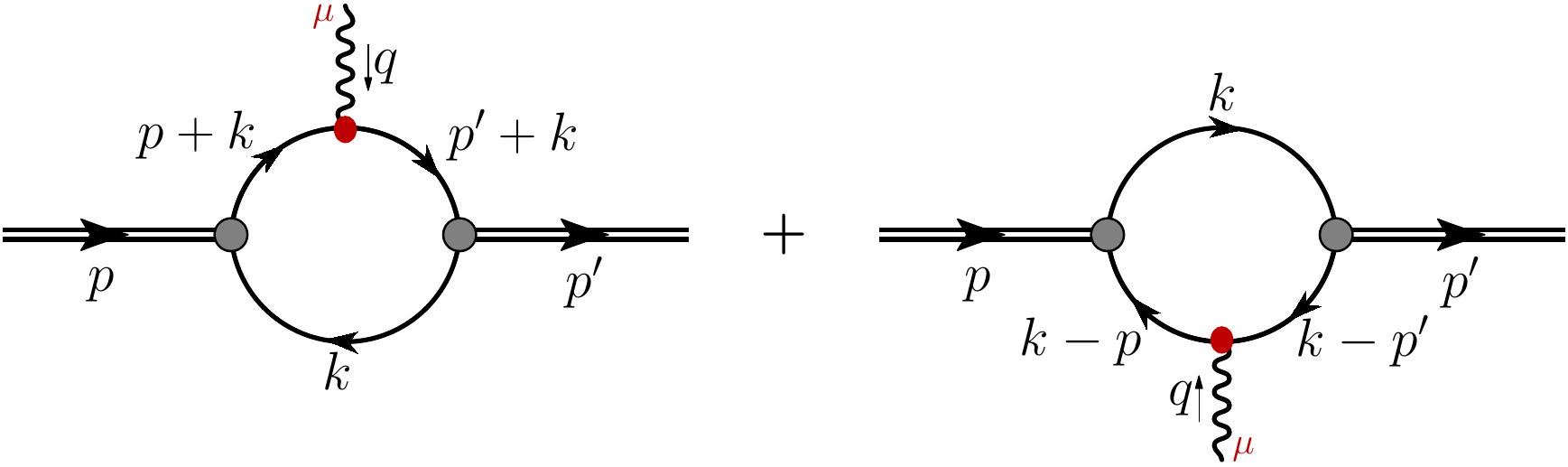}
\caption{(Color online) Feynman diagrams representing the electromagnetic current of the pion or kaon.} 
\label{fig:emvertex1}
\end{figure}

The matrix element of the electromagnetic current for a pseudoscalar meson $\a$ is characterized by a single form factor:
\begin{align}
\label{eq:formfactor1}
J_\a^{\mu} (p',p) &= \lf(p'^\mu + p^\mu\rg) F_\a (Q^2),
\end{align}
where $p^\mu$ is the initial and $p'^\mu$ the final hadron momentum. In the NJL model the form factor of a pseudoscalar meson is given by the sum of the two Feynman diagrams illustrated in Fig.~\ref{fig:emvertex1}, which read
\begin{align}
\label{eq:j1}
j^{\mu}_{1,\a}\lf(p',p\rg) &= i\,Z_\a \int \frac{d^4k}{(2\pi)^4} \
\mathrm{Tr}\Big[\g_5\l_\a^\dagger\,S(p'+k) \no \\
&\hs*{1mm}
\times \L^\mu_{\g Q}(p'+k,p+k)\,S(p+k)\,\g_5\l_\a S(k)\Big], \\
\label{eq:j2}
j^{\mu}_{2,\a}\lf(p',p\rg) &= i\,Z_\a \int \frac{d^4k}{(2\pi)^4} \
\mathrm{Tr}\Big[\g_5\l_\a\,S(k-p) \no \\
&
\times \L^\mu_{\g Q}(k-p,k-p')\,S(k-p')\,\g_5\l_\a^\dagger S(k)\Big],
\end{align}
where the trace is over Dirac, color and flavor indices, $S(p)$ is the quark propagator and $\L^\mu_{\g Q}(p',p)$ the quark-photon vertex, both in flavor space. These expressions are valid for all $\a=\pi^{\pm},\,\pi^0,\,K^{\pm},\,K^0,\bar{K}^0$. Evaluating these expressions we find that the pseudoscalar form factors of interest are given by
\begin{align}
\label{eq:fullpi}
F_{\pi^{+}}(Q^2) &= F_{1U}(Q^2) f^{ud}_{\pi^+}(Q^2) - F_{1D}(Q^2) f^{du}_{\pi^+}(Q^2), \\
F_{K^{+}}(Q^2) &= F_{1U}(Q^2)\,f^{us}_{K^+}(Q^2) - F_{1S}(Q^2)\,f^{su}_{K^+}(Q^2), \\
\label{eq:fullK0}
F_{K^0}(Q^2) &= F_{1D}(Q^2)\,f^{ds}_{K^0}(Q^2) - F_{1S}(Q^2)\,f^{sd}_{K^0}(Q^2),
\end{align}
where the $F_{1Q}(Q^2)$ are defined in Eq.~\eqref{eq:chargeffs}. The universal body form factor reads~\cite{Hutauruk:2016sug}:
\begin{align}
&f^{ab}_\a(Q^2) = \frac{3\,Z_\a}{4\,\pi^2} \int_0^1 dx \int \frac{d\tau}{\tau}\ e^{-\tau\lf[M_a^2 + x(1-x)\,Q^2\rg]} \no \\
&+ \frac{3\,Z_\a}{4\,\pi^2}  \int_0^1\! dx\! \int_0^{1-x}\! dz\! \int\! d\tau\, \no \\
&\quad \times
\Bigl[(x+z)\lf[m_\a^2 + (M_a - M_b)^2\rg] + 2\,M_b\lf(M_a - M_b\rg)\Bigr] \no \\
&
\quad \times
e^{-\tau\lf[(x+z)(x+z-1)\,m_\a^2 + (x+z)\,M_a^2 + (1-x-z)\,M_b^2 + x\,z\,Q^2\rg]},
\label{eq:bodyff}
\end{align}
where the first superscript ($a$) on the body form factor indicates the struck quark and the second ($b$) the spectator. Form factors for $\a=\pi^{-},\,\pi^0,\,K^{-},\,\bar{K}^0$ can straightforwardly be determined from Eqs.~\eqref{eq:fullpi}--\eqref{eq:fullK0} by the appropriate substitution of quark flavor, meson mass and Bethe-Salpeter vertex normalization.

\begin{figure}[tbp]
\centering\includegraphics[width=\columnwidth]{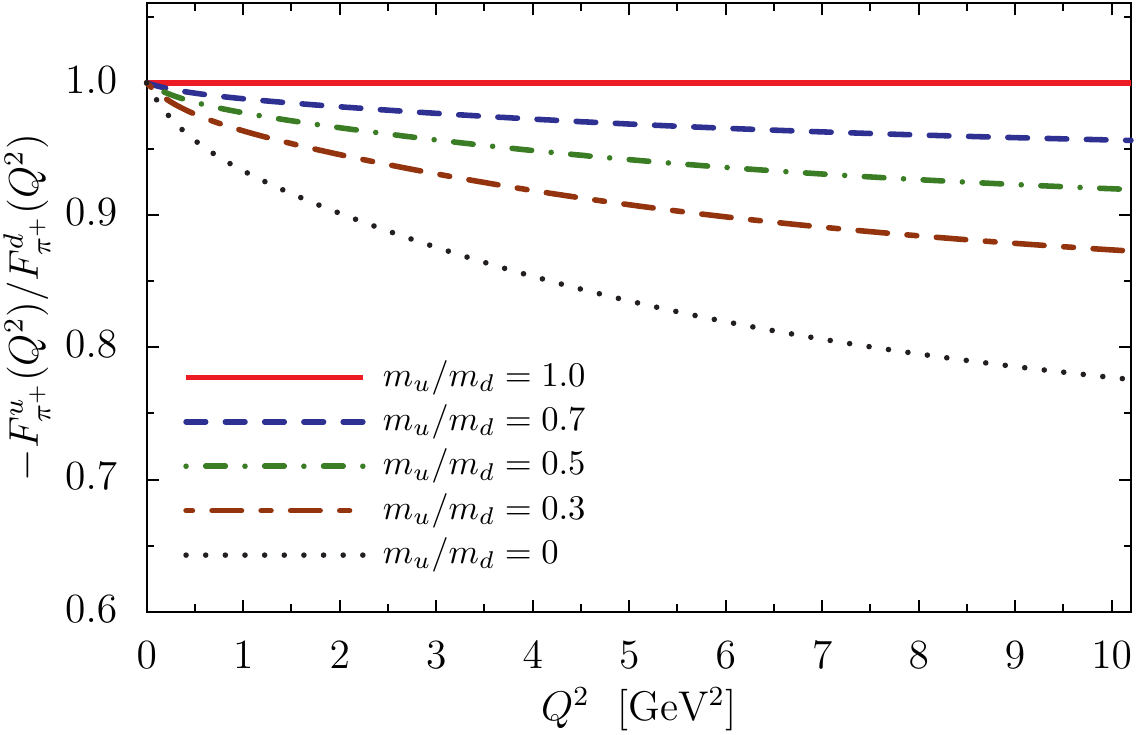} \\[0.8em]
\centering\includegraphics[width=\columnwidth]{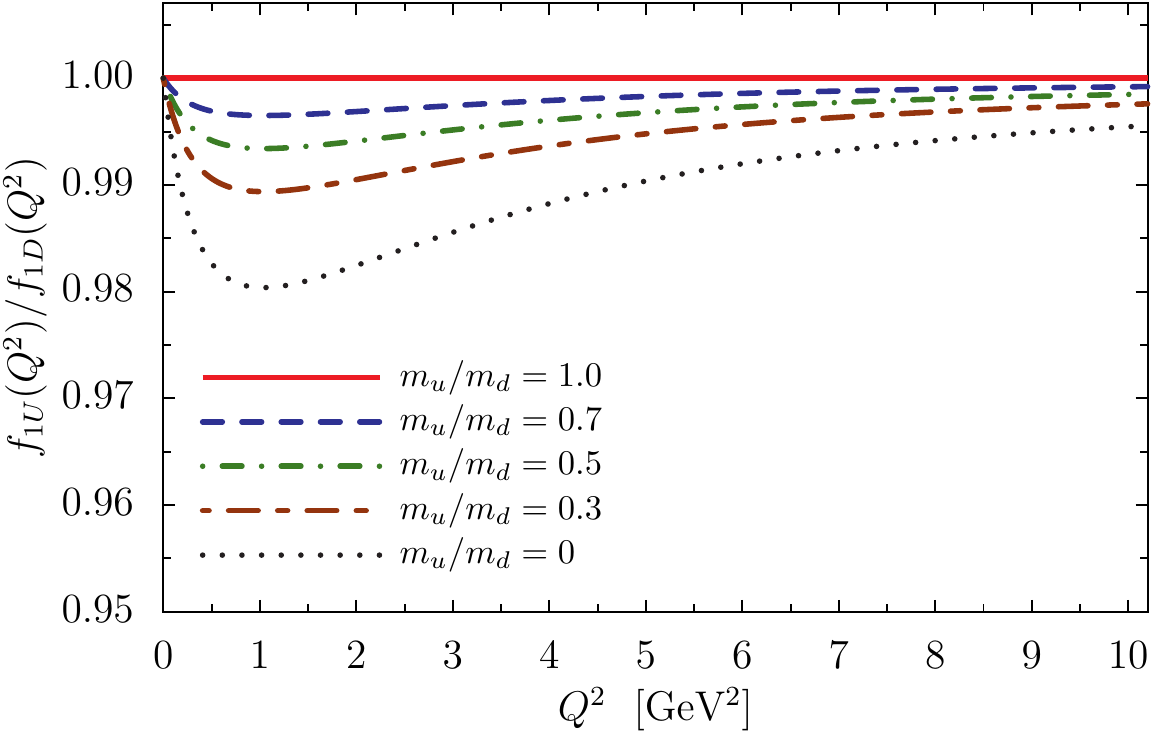} 
\caption{(Color online) {\it Upper panel:} Ratio of the $u$ and $d$ quark sector form factors in the $\pi^+$ for various values of $m_u/m_d$. {\it Lower panel:} CSB effects in the dressed $u$ and $d$ quark-photon vertex, where the functions $f_{1U,1D}$ are the quark-sector dressed quark charge form factors defined in Eq.~\eqref{eq:chargeffs}.}
\label{fig:csvformfactor1}
\end{figure}

As the first example of CSB we compare the $u$ quark sector form factor of the $\pi^+$, $F^u_{\pi^+} (Q^2)$, with the corresponding $d$ quark sector form factor, $F^{d}_{\pi^+}(Q^2)$, where the quark-sector form factors are defined by
\begin{align}
F_\a(Q^2) = e_u\,F_\a^u(Q^2) + e_d\,F_\a^d(Q^2) + e_s\,F_\a^s(Q^2) + \ldots
\label{eq:quarksector}
\end{align}
With this definition and Eqs.~\eqref{eq:fullpi}--\eqref{eq:fullK0} it is straightforward to obtain the quark-sector form factors. Our results for the ratio $F^{u}_{\pi^+}(Q^2)/F^{d}_{\pi^+}(Q^2)$ at various values of $m_u/m_d$ are shown in the upper panel of Fig.~\ref{fig:csvformfactor1}. We find that this ratio decreases from unity as $m_u/m_d$ gets smaller, which reflects that the $u$ quark sector charge radius is larger in magnitude than the $d$ quark sector. This result has a natural physics interpretation, because when $m_u/m_d < 1$ we have $M_u < M_d$ and the lighter dressed $u$-quark has a larger probability to be further from the charge center of the $\pi^+$. Results for the quark-sector charge radii in the $\pi^+$ are given in Tab.~\ref{tab:radii}, where for a realistic value of $m_u/m_d \simeq 0.5$ we find CSB effects of the size $[|r^u_{\pi^+}| - |r^d_{\pi^+}|]/[|r^u_{\pi^+}| + |r^d_{\pi^+}|] \simeq 0.7$\%. Such effects are unlikely to be measurable in the foreseeable future, however with increasing $Q^2$ we find that CSB effects  increase substantially, reaching about 8\% at $Q^2 \simeq 10\,$GeV$^2$ for realistic values of $m_u/m_d$. 

\begin{figure}[tbp]
\centering
\includegraphics[width=\columnwidth]{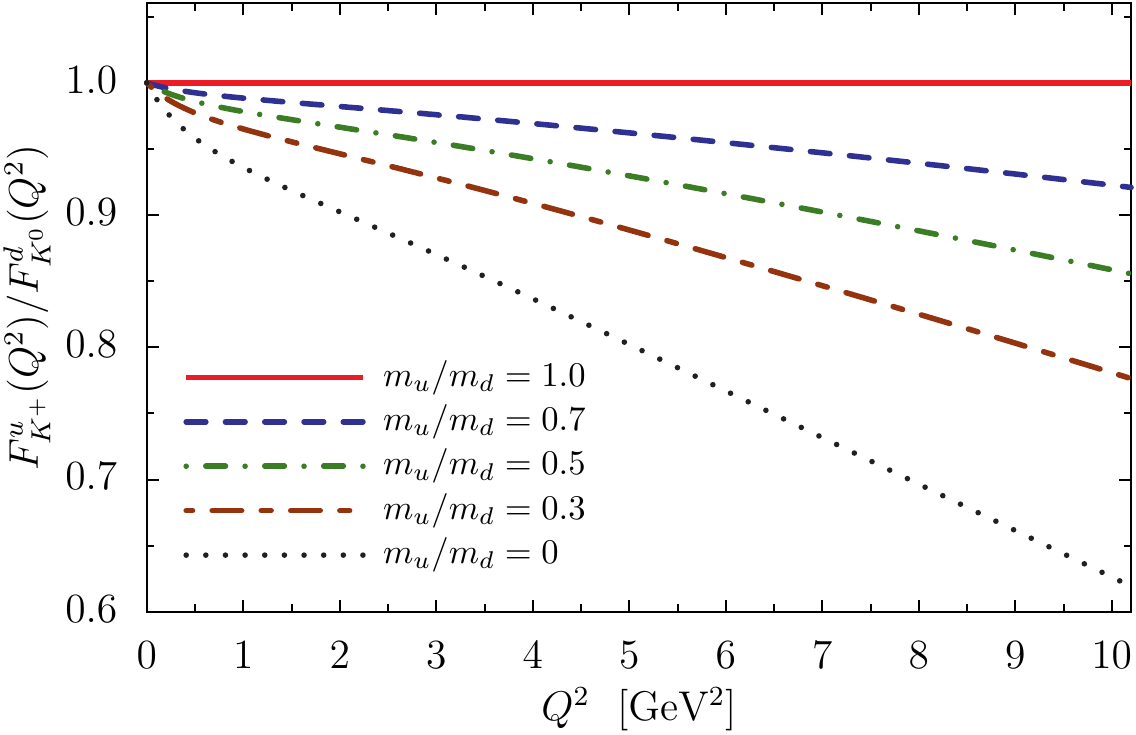} \\[0.8em]
\includegraphics[width=\columnwidth]{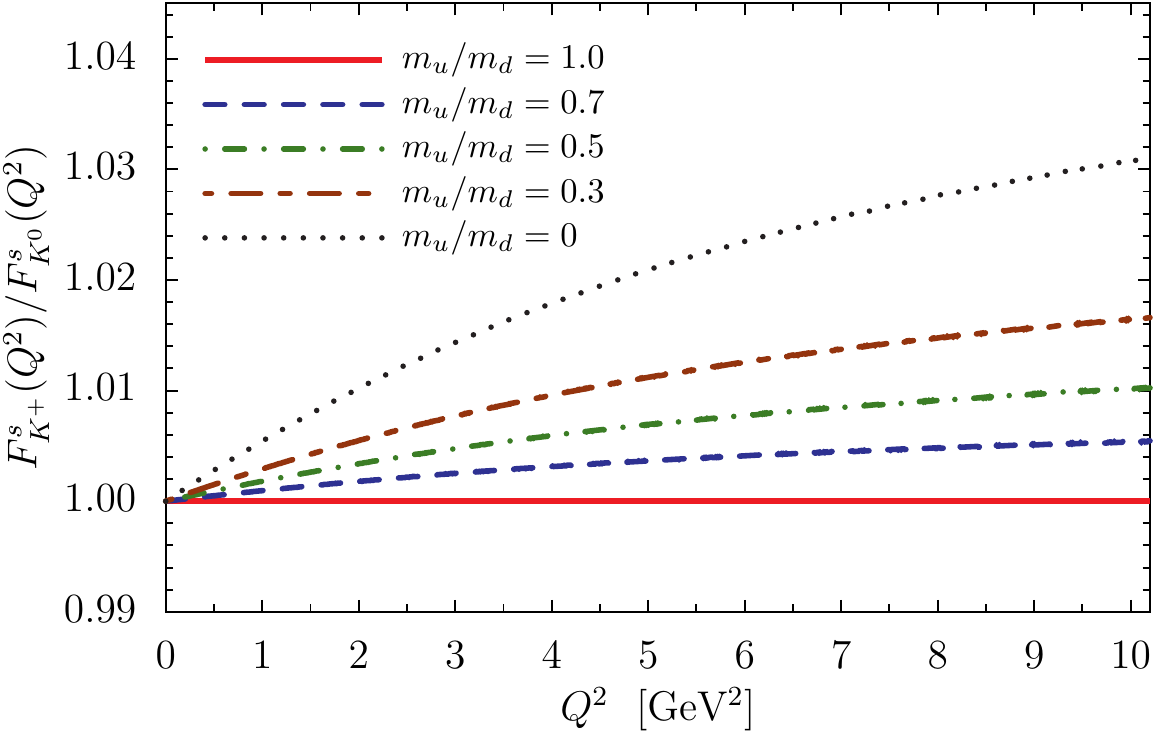}
\caption{(Color online) {\it Upper panel:} Results for CSB effects between the $u$ quark sector form factors in the $K^+$ and the $d$ quark sector form factors in the $K^0$. These CSB effects are found to be about twice that of the pion. {\it Lower panel:} Comparison between the $s$ quark sectors in $K^+$ and the $K^0$, which is a measure of the environment sensitivity for the $s$ quark in both mesons. These effects are an order of magnitude smaller that the CSB effects.}
\label{fig:kaonff}
\end{figure}

This interesting result is traced to the body form factors given in Eq.~\eqref{eq:bodyff}, because CSB effects in the quark-photon vertex are small and vanish for increasing $Q^2$, as illustrated in the lower panel of Fig.~\ref{fig:csvformfactor1}. At large $Q^2$ the leading CSB piece of the body form factors given in  Eq.~\eqref{eq:bodyff} behaves as
\begin{align}
Q^2\,f^{ab}_{\a,\text{CSB}}(Q^2)\ &\stackrel{Q^2 \gg m_\a^2}{\propto} \no \\
&\hs*{-8mm}
\delta M\,M
\int \frac{d\tau}{\tau}\,e^{-\tau\,M^2}\lf[\g_E -1 + \text{log}\lf(Q^2\,\tau\rg)\rg],
\label{eq:largeQ2}
\end{align}
where $M = \frac{1}{2}\lf(M_a + M_b\rg)$ and $\delta M = \frac{1}{2}\lf(M_a - M_b\rg)$. Therefore we find that CSB effects grow logarithmically with $Q^2$. In the asymptotic limit ($Q^2 \to \infty$) of QCD the pion's electromagnetic form factor is predicted to behave as~\cite{Farrar:1979aw,Lepage:1979zb,Lepage:1980fj}
\begin{align}
Q^2\,F_{\pi^+}(Q^2) &\stackrel{Q^2 \to \infty}{=} \frac{16\,\pi}{3}\,f_{\pi^+}^2\,\a_s(Q^2)\, \mathpzc{w}_{\pi^+}^2(Q^2),
\end{align}
where $\a_s(Q^2)$ is the strong running coupling, $\mathpzc{w}_{\pi^+} = \int_0^1 \frac{dx}{x}\ \varphi_{\pi^+}(x,Q^2)$, and $\varphi_{\pi^+}(x,Q^2)$ is the pion's leading distribution amplitude, which can be expressed as
\begin{align}
\hs*{-2.5mm}\varphi_{\pi^+}(x,Q^2) &= 6\,x\,(1-x) \no \\
&\hs*{1mm}
\times
\bigg[1 +\!\!\! \sum_{n=2,\,4,\ldots}\!\!\! a_n^{\pi^+}(m_u,m_d,Q^2)\,C_n^{3/2}(2x-1)\bigg].
\end{align}
The expansion is in Gegenbauer--$3/2$ polynomials, and the coefficient functions $a_n(Q^2)$ have a quark mass dependence but vanish logarithmically as $Q^2 \to \infty$. Therefore QCD predicts that at scales where $a_n \simeq 0$ CSB effects in the quark-sector $\pi^+$ electromagnetic form factor must be negligible. However, a Dyson-Schwinger equation study~\cite{Chang:2013nia} and an analysis of lattice QCD results~\cite{Cloet:2013tta} demonstrates that this condition is only satisfied at multi-TeV scales. Therefore, we predict that CSB effects from the $u$ and $d$ quark mass difference should initially increase with $Q^2$, then at scales $Q \gg \L_{\rm QCD}$ when perturbative QCD effects start to dominate, they should begin to decrease and then vanish in the asymptotic limit.

\begin{table}[bp]
\addtolength{\tabcolsep}{3.0pt}
\addtolength{\extrarowheight}{2.2pt}
\begin{tabular}{cccccccccccccccccccc}
\hline\hline
$m_u/m_d$ & $r^u_{\pi^+}$ & $r^d_{\pi^+}$ & $r^u_{K^+}$ & $r^d_{K^0}$ & $r^s_{K^+}$ & $r^s_{K^0}$ \\[0.2em]
\hline
0        & 0.634 & $-$0.608 & 0.650 & 0.625 & $-$0.436 & $-$0.438 \\
0.1      & 0.632 & $-$0.610 & 0.647 & 0.627 & $-$0.436 & $-$0.438 \\
0.3      & 0.628 & $-$0.614 & 0.644 & 0.631 & $-$0.437 & $-$0.438  \\
0.5      & 0.625 & $-$0.616 & 0.641 & 0.633 & $-$0.437 & $-$0.437 \\
0.7      & 0.623 & $-$0.618 & 0.639 & 0.635 & $-$0.437 & $-$0.437 \\
0.9      & 0.621 & $-$0.620 & 0.638 & 0.636 & $-$0.437 & $-$0.437 \\
1        & 0.621 & $-$0.621 & 0.637 & 0.637 & $-$0.437 & $-$0.437  \\
\hline\hline
\end{tabular}
\caption{Results for the quark sector radii in the $\pi^+$, $K^+$ and $K^0$ given in units of fm. The radii are defined by 
$r = {\rm sign}\lf(\big<r^2\big>\rg)\,\sqrt{|\big<r^2\big>|}$, where $\big<r^2\big> = -\frac{6}{F(0)}\ \pl F(Q^2)/\pl Q^2\Big|_{Q^2=0}$.}
\label{tab:radii}
\end{table}

In the upper panel of Fig.~\ref{fig:kaonff} we illustrate CSB effects between the $u$ quark sector form factor in the $K^+$ and the $d$ quark sector form factor in the $K^0$. We find that the ratio $F^u_{K^+}(Q^2)/F^d_{K^0}(Q^2)$ is smaller than unity, and that the CSB effects grow with increasing $Q^2$, where for the kaon these effects are about twice that of the pion for large $Q^2$. We therefore find that the $u$ quark charge radius in the $K^+$ is larger in magnitude than the $d$ quark radius in the $K^0$, which is in agreement with expectation from the fact that $M_u < M_d$. For $m_u/m_d \simeq 0.5$ we find CSB effects in the quark sector radii of $[|r^u_{K^+}| - |r^d_{K^0}|]/[|r^u_{K^+}| + |r^d_{K^0}|] \simeq 0.6$\% which is similar to that found in the pion.  These results are summarized in Tab.~\ref{tab:radii}. 

As $Q^2$ increases the CSB effects largely result from the body form factors, not the dressing of the quark-photon vertex, which vanishes at large $Q^2$. In Eq.~\eqref{eq:bodyff} for the body form factors, the dominant CSB effect between the quark-sectors in the $\pi^+$ comes from the term linear in the mass difference, that is, $2\,M_b\,(M_a-M_b)$. In the kaon however, there are two sources of CSB, one directly from the quark mass difference $\d M = \frac{1}{2}\lf(M_u-M_d\rg)$ and the other from the mass difference between the kaons $\d m_K^2 = \frac{1}{2}\lf(m_{K^+}^2 - m_{K^0}^2\rg) \simeq \lf(m_{K^+} + m_{K^0}\rg)\,\d M$. These CSB effects enter with the same sign, which explains why CSB effects in the kaon sector are larger than in the charged pion.
Again, in the asymptotic limit of QCD these effects will vanish, however at all current and foreseeable facilities CSB may remain large over accessible energy scales because the quark mass dependent terms in the kaon's distribution amplitude, $a_n(Q^2)$, only become negligible at multi-TeV scales. 

In the lower panel of Fig.~\ref{fig:kaonff} we illustrate the ratio $F^s_{K^+}(Q^2)/F^s_{K^0}(Q^2)$ for various values of $m_u/m_d$. We find that this ratio is larger than unity, which implies that the $s$-quark charge radius in the $K^+$ is smaller in magnitude than the same radius in the $K^0$ (see Table~\ref{tab:parameters}). This is consistent with a simple picture for the kaon, where the lighter $u$ quark is less able to pull the heavier $s$ quark away from the charge center of kaon. We note however, that these environment sensitivity effects are at the few percent level, and therefore much smaller than the CSB effects. 

\begin{figure}[tbp]
\centering\includegraphics[width=\columnwidth]{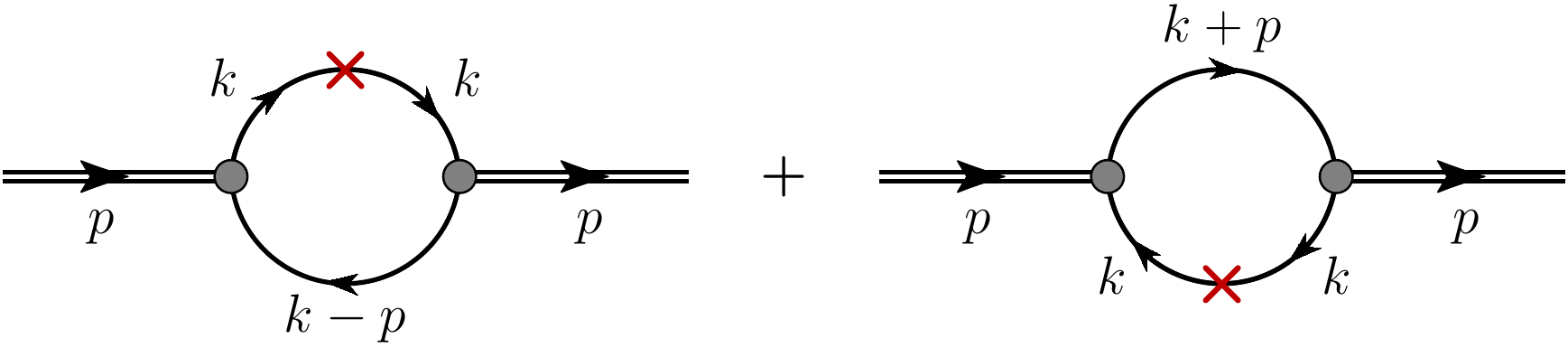}
\caption{\label{fig:strucfun1} (Color online) Feynman diagrams representing the quark distribution functions in the pion or kaon. The operator insertion has the form $\g^+ \delta \lf(k^+ - x\,p^+\rg)\hat{P}_q$, where $\hat{P}_q$ is the projection operator for quarks of flavor $q$.}
\end{figure}

Another key set of observables where CSB effects may play an important role are the quark distribution functions of hadrons, where the pion, kaon and nucleon are of particular interest. The leading-twist quark distributions in a hadron $\a$ are defined by the matrix element~\cite{Barone:2001sp}
\begin{align}
\label{eq:valence1}
q_\a(x) &= \int \frac{d\xi^{-}}{4\pi}\ e^{ix\,p^{+}\,\xi^{-}}\
\langle \a\,| \bar{\psi}_{q}(0) \gamma^{+} \psi_q (\xi^{-})|\,\a \rangle_{c},
\end{align}
where $x =  \frac{k^+}{p^+}$ is the lightcone momentum fraction of the struck quark, with light-cone momentum $k^+$, relative to the parent hadron, with light-cone momentum $p^+$, $q$ labels the quark flavor, and the subscript $c$ denotes a connected matrix element. Here we focus on the pion and kaon PDFs, where from Eq.~\eqref{eq:valence1} one may readily show~\cite{Barone:2001sp} that the PDFs of the pion or kaon in the NJL model are given by the two Feynman diagrams in Fig.~\ref{fig:strucfun1}. The operator insertion for a quark distribution of flavor $q$ is $\g^+\delta\lf(p^+x - k^+\rg)\hat{P}_q$, where the quark-flavor 
projection operators read $\hat{P}_{u/d} = \frac{1}{2}\Big(\frac{2}{3}\,\ident \pm \l_3 + \frac{1}{\sqrt{3}}\,\l_8\Big)$ and $\hat{P}_s = \frac{1}{3}\,\ident - \frac{1}{\sqrt{3}}\,\l_8 $. Using the relation $\bar{q}(x) = -q(-x)$, the valence quark and 
anti-quark distributions in the pion or kaon are given by
\begin{align}
\label{eq:valence3}
q_\a(x) &= \ph{-}i\,\frac{Z_\a}{2} \int \frac{d^4k}{(2\pi)^4}\ \delta\lf(p^+x - k^+\rg) \no \\
&\hs{5mm}
\times \mathrm{Tr}\lf[\gamma_5\l_\a^\dagger\,S(k)\,\g^+\hat{P}_q\,S(k)\,\gamma_5\l_\a\,S(k-p)\rg], \allowdisplaybreaks \\
\label{eq:valence4}
\bar{q}_\a(x) &= -i\,\frac{Z_\a}{2} \int \frac{d^4k}{(2\pi)^4}\ \delta\lf(p^+x + k^+\rg) \no \\
&\hs{5mm}
\times \mathrm{Tr}\lf[\gamma_5\l_\a\,S(k)\,\g^+\hat{P}_q\,S(k)\,\gamma_5\l_\a^\dagger\,S(k+p)\rg].
\end{align}
where, as for the form factors, $\a=\pi^{\pm},\,\pi^0,\,K^{\pm},\,K^0,\bar{K}^0$ and $\l_\a$ are the appropriate flavor matrices which we list below Eq.~\eqref{eq:bubblegraphtot}.

To determine the valence quark distributions from Eqs.~\eqref{eq:valence3}--\eqref{eq:valence4} we first take the moments, defined by $\mathcal{A}_n = \int_0^1 dx\, x^{n-1}\, q(x)$ where $n = 1,\,2,\ldots$, which removes the delta function. Then, using Feynman parametrization and standard manipulations of loop integrals, we can again express these moments in the form given for $\mathcal{A}_n$, where the integral over $x$ originates from the Feynman parametrization. One can then simply read off the expression for the quark distributions, which for the $\pi^+$ in the proper-time regularization scheme are:
\begin{align}
\label{eq:valence5}
u_{\pi^+}(x)  &= \frac{3\,Z_{\pi^+}}{4\,\pi^2} \int d\tau\
e^{-\tau\lf[x(x - 1)\,m_{\pi^+}^2 + x\,M_d^2 + (1-x)\,M_u^2\rg]} \no \\
&\hs{8mm}
\times \left[\frac{1}{\tau} + x(1 - x)\left[m_{\pi^+}^2 - (M_d - M_u)^2\right] \right], \allowdisplaybreaks\\ 
\label{eq:valence6}
\bar{d}_{\pi^+}(x)  &= \frac{3\,Z_{\pi^+}}{4\,\pi^2} \int d\tau\
e^{-\tau\lf[x(x - 1)\,m_{\pi^+}^2 + x\,M_u^2 +  (1-x)\,M_d^2\rg]} \no \\
&\hs{8mm}
\times \left[\frac{1}{\tau} + x(1 - x)\left[m_{\pi^+}^2 - (M_d - M_u)^2\right]\right].
\end{align}
From these expressions it is straightforward to also obtain the PDFs in the $\pi^0$, $\pi^-$ and the kaons, by using the appropriate Bethe-Salpeter vertex normalization $Z_\a$ and hadron mass $m_\a$, and making the necessary substitutions for the dressed quark masses. For example, by making the substitution $M_d \to M_s$, the $u$-quark distribution in the $K^+$ is obtained from Eq.~\eqref{eq:valence5} and the $\bar{s}$-quark distribution in the $K^+$ is obtained from Eq.~\eqref{eq:valence6}. For each quark distribution we find that the baryon number and momentum sum rules are satisfied exactly.

\begin{figure}[tbp]
\centering\includegraphics[width=\columnwidth]{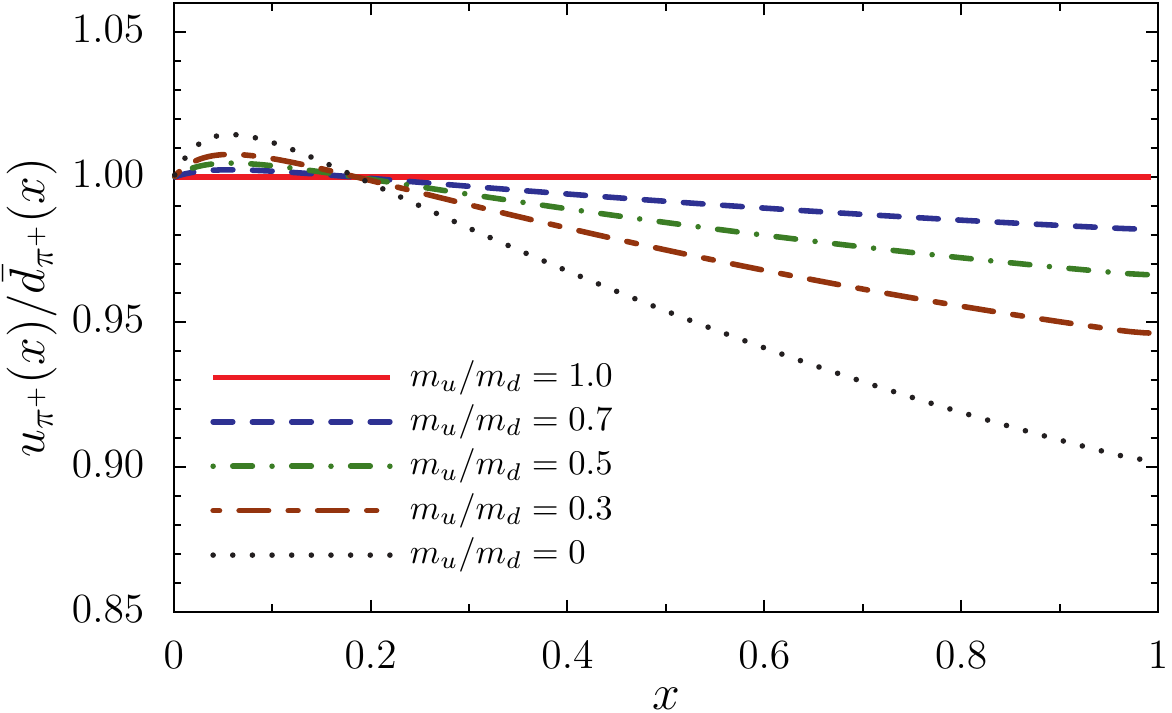} \\[0.8em]
\centering\includegraphics[width=\columnwidth]{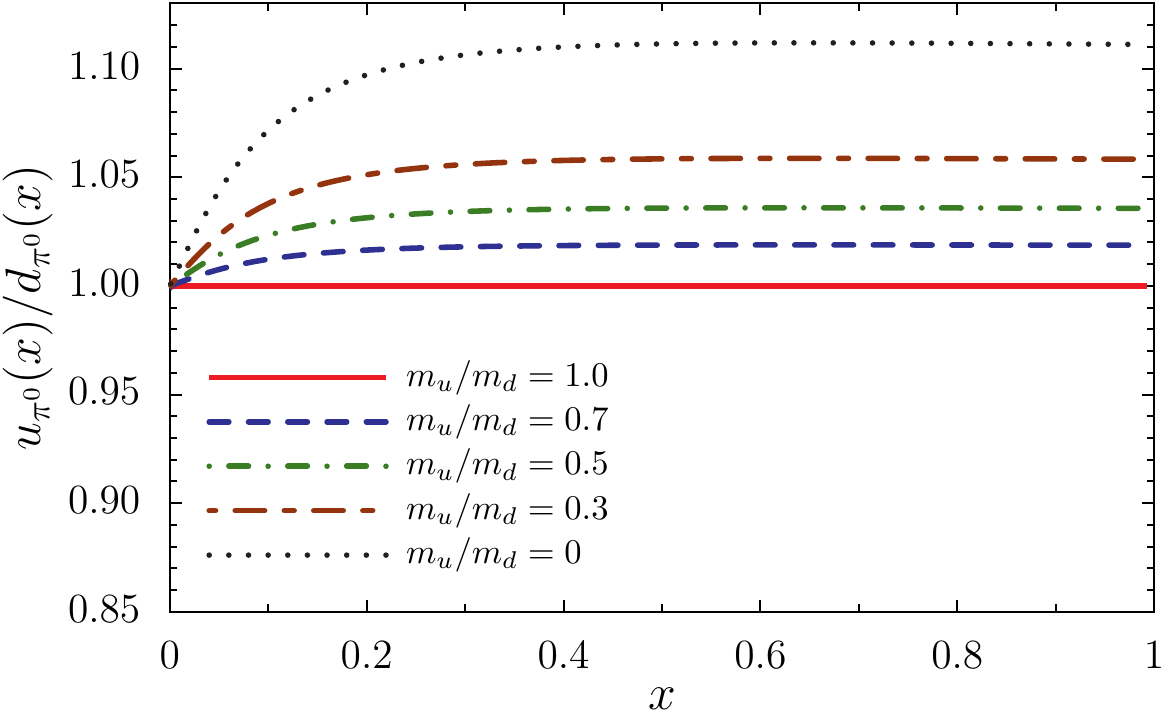} 
\caption{(Color online) {\it Upper panel:} Ratio of the $u$ quark distribution to the $\bar{d}$ quark distribution in the $\pi^+$, after QCD evolution to a scale of $Q^2 = 5\,$GeV$^2$, for various values of current quark mass ratio $m_u/m_d$.  {\it Lower panel:} Ratio of the $u$ quark distribution to the $d$ quark distribution in the neutral pion, at a scale of $Q^2 = 5\,$GeV$^2$.}
\label{fig:csvpdf2}
\end{figure}

Results for the CSB effects in the $\pi^+$, as expressed through the ratio $u_{\pi^+}(x)/\bar{d}_{\pi^+}(x)$, are presented in the upper panel of Fig.~\ref{fig:csvpdf2}. These results have been evolved~\cite{Bertone:2013vaa} from the model scale of $Q_0^2 = 0.16\,$GeV$^2$, which was determined in previous work~\cite{Cloet:2005pp,Cloet:2007em}, to the scale $Q^2 = 5\,$GeV$^2$, where we are plotting results for the quark distributions (not the valence quark distributions). For $x \gtrsim 0.2$ we find that this ratio is less than unity in agreement with the expectation that the lighter $u$-quark should carry less light-cone momentum than the heavier $d$-quark. For $x \simeq 0.2$ this ratio crosses unity, where the position is largely independent of CSB effects, but is $Q^2$ dependent and in each case has its origin in the need to satisfy the baryon number and momentum sum rules. For $x \lesssim 0.2$ we find that CSB effects are suppressed by the Dokshitzer-Gribov-Lipatov-Altarelli-Parisi (DGLAP) evolution which usually treats the light quarks as massless, because mass effects are suppressed by $1/Q^2$. Although not plotted, we also investigated CSB effects from QED evolution using the code from Ref.~\cite{Bertone:2013vaa}, and found only slightly larger effects of the order of 1-2\% and mainly concentrated at very large $x$. 

In the lower panel of Fig.~\ref{fig:csvpdf2} CSB effects in the PDFs of the neutral pion are presented. Here, we find that the ratio $u_{\pi^0}(x)/d_{\pi^0}(x)$ is always greater than unity when CSB effects are included. In contrast to the $\pi^+$, this implies that the lighter $u$-quark carries more lightcone momentum in the $\pi^0$ than the heavier $d$-quark. The simple reason for this is that for $m_u < m_d$ the $\bar{u}u$ component of the $\pi^0$ is more likely than the $\bar{d}d$ component, where in our model the probability of each component is the same as its lightcone momentum fraction.\footnote{Note, when $m_u \neq m_d$ the $\pi^0$ mixes with the $\eta$ and $\eta'$, however here we ignore these mixing effects which are unlikely to change our findings.} Again the ratios vanish become unity at small $x$ because of DGLAP evolution, and for the $\pi^0$ a crossing of the unity line is not required because of baryon number conservation. Finally, in general we find that the CSB effects in the PDFs are much smaller than in the electromagnetic form factors at high momentum transfer.

\begin{figure}[tbp]
\centering\includegraphics[width=\columnwidth]{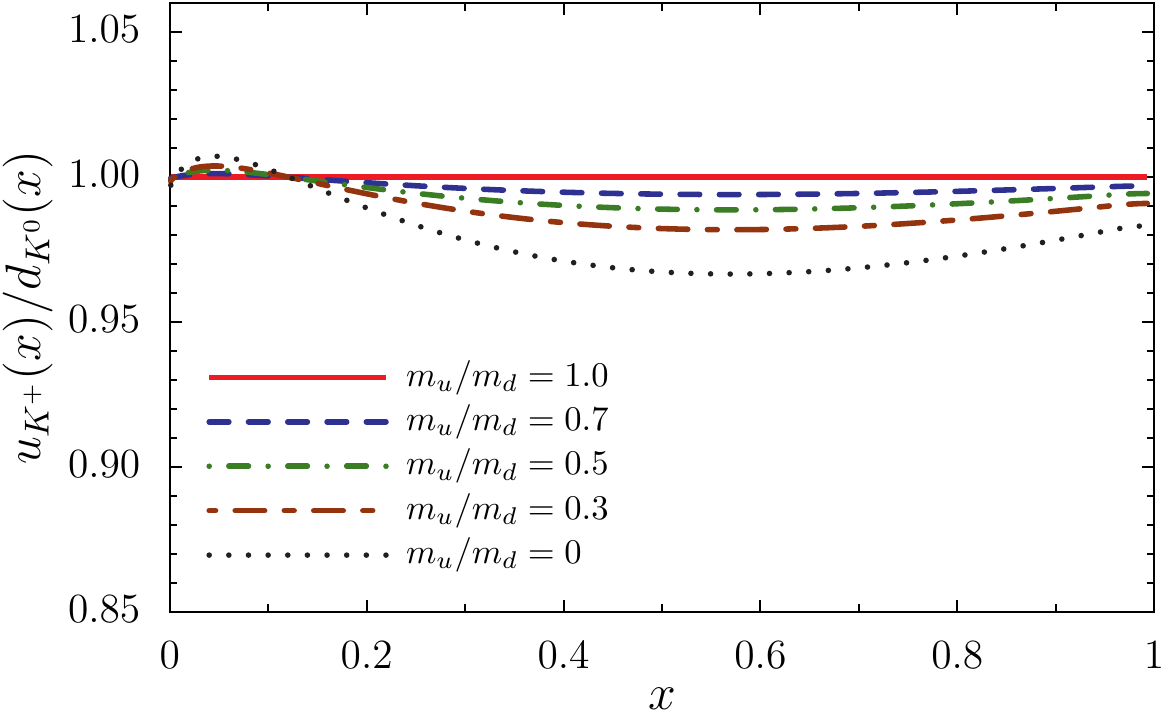} \\[0.8em]
\centering\includegraphics[width=\columnwidth]{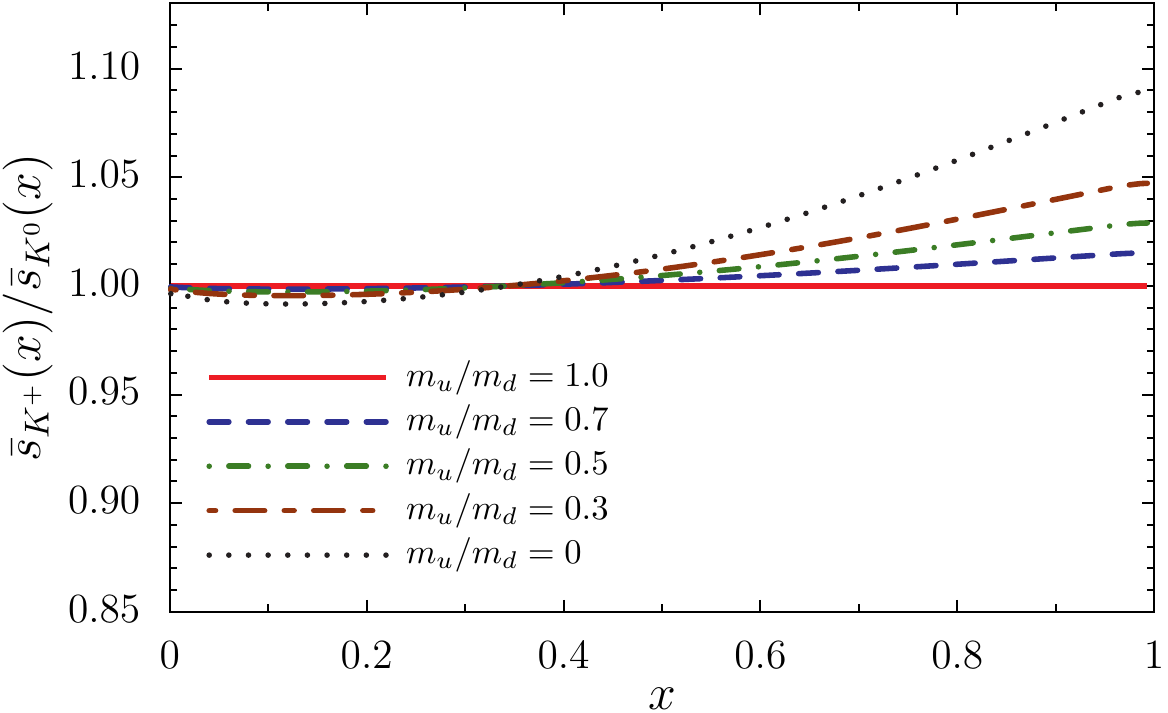} 
\caption{(Color online) \label{fig:csvpdfkaon2} Ratio of the $u$ quark distribution in the $K^+$ to the $d$ quark distribution in the $K^0$, after QCD evolution to a scale of $Q^2 = 5\,$GeV$^2$, for various values of current quark mass ratio $m_u/m_d$.  {\it Lower panel:} Ratio of the $\bar{s}$ quark distribution in the $K^+$ to the same PDF in the $K^0$ at a scale of $Q^2 = 5\,$GeV$^2$. This ratio is a measure of environment sensitivity effects.}
\end{figure}

\begin{table}[bp]
\addtolength{\tabcolsep}{7.0pt}
\addtolength{\extrarowheight}{2.2pt}
\begin{tabular}{cccccccccccccccccccc}
\hline\hline
$m_u/m_d$ & $\big<x\,\d q_{\pi^+}\big>$ & $\big<\d q_{\pi^0}\big>$ & $\big<x\,\d q_{\pi^0}\big>$ & $\big<x\,\d q_{K}\big>$ \\[0.2em]
\hline
0        & 0.0174 & $-$0.0532 & $-$0.0266 & 0.0086 \\
0.1      & 0.0143 & $-$0.0435 & $-$0.0218 & 0.0070 \\
0.3      & 0.0094 & $-$0.0286 & $-$0.0143 & 0.0046 \\
0.5      & 0.0058 & $-$0.0177 & $-$0.0089 & 0.0029 \\
0.7      & 0.0031 & $-$0.0094 & $-$0.0047 & 0.0015 \\
0.9      & 0.0009 & $-$0.0028 & $-$0.0014 & 0.0005 \\
1        & 0         & 0          &  0         & 0    \\
\hline\hline
\end{tabular}
\caption{Results for moments of the quantities: $\delta q_{\pi^+} (x) = \bar{d}_{\pi^+}(x) - u_{\pi^+}(x)$, $\delta q_{\pi^0} (x) = d_{\pi^0}(x) - u_{\pi^0}(x)$ and  $\delta q_K (x) = d_{K^0}(x) - u_{K^+}(x)$. These results are at the model scale of $Q^2 = 0.16\,$GeV$^2$, where there are no sea quarks, so the first moments of $\delta q_{\pi^+} (x)$ and $\delta q_K (x)$ must vanish, and are therefore not tabulated.}
\label{tab:moments}
\end{table}

In the upper panel of Fig.~\ref{fig:csvpdfkaon2} we illustrate CSB effects which cause differences between the $u$-quark PDF in the $K^+$ and the $d$-quark PDF in the $K^0$. At a scale of $Q^2 = 5\,$GeV$^2$, we find that these CSB effects are at the few percent level, making CSB effects in the kaon PDFs much smaller than these effects within the pion. Interestingly, this is the opposite of what we found for CSB effects between the pion and kaon electromagnetic form factors. The reason for the difference can be understood by examining Eqs.~\eqref{eq:valence5}--\eqref{eq:valence6}, and their analogs for the $K^+$ and $K^0$. In analogy with the associated CSB effects in the pion and kaon electromagnetic form factors, the charged pion PDFs receive CSB from the quark mass difference $\d M = \frac{1}{2}\lf(M_u-M_d\rg)$, however in the kaon sector CSB from $\d M$ also induces CSB in the kaon masses $\d m_K^2 = \frac{1}{2}\lf(m_{K^+}^2 - m_{K^0}^2\rg) \simeq \lf(m_{K^+} + m_{K^0}\rg)\,\d M$, and for the kaon PDFs these two CSB contributions have opposite sign making the effect smaller than in the pion PDFs. Note, the opposite was found for the CSB in the pion and kaon electromagnetic form factors. In the lower panel of Fig.~\ref{fig:csvpdfkaon2} we show results for the ratio $\bar{s}_{K^+}(x)/\bar{s}_{K^0}(x)$, which is a measure of environment sensitivity. For realistic values of $m_u/m_d$ we find effects at the few percent level, that are greater than unity and maximal when $x \to 1$, which is consistent with the expectation that the $\bar{s}$-quark in the $K^+$ should carry more lightcone momentum than the $\bar{s}$-quark in the $K^0$ because $M_u < M_d$.

As a final measure of CSB effects in the pion and kaon PDFs we consider the quantities: $\delta q_{\pi^+} (x) = \bar{d}_{\pi^+}(x) - u_{\pi^+}(x)$, $\delta q_{\pi^0} (x) = d_{\pi^0}(x) - u_{\pi^0}(x)$, and  $\delta q_K (x) = d_{K^0}(x) - u_{K^+}(x)$. Considering only valence distributions, the first moments of $\delta q_{\pi^+} (x)$ and $\delta q_K (x)$ must vanish because of baryon number conservation, however higher moments $\big<x^{n-1} \d q\big>$ of these quantities and all moments of $\delta q_{\pi^0} (x)$ need not vanish. In Table~\ref{tab:moments} we give results for these moments for various values of $m_u/m_d$, at the model scale. These results again demonstrate that CSB effects in the pion and kaon PDFs are typically at the few percent level, with CSB in the kaon sector about half the size as for the charged pion, and CSB within the neutral pion much larger than in the charged pion.

\section{SUMMARY\label{sec:summary}}
In summary, we have investigated CSB effects in the spacelike electromagnetic form factors and quark distribution functions of the pion and kaon using the NJL model with the proper-time regularization scheme. We found that the effect of CSB arising from the light quark mass differences is surprisingly large in the quark-sector elastic form factors at large momentum transfer. This is especially dramatic in the kaon, where for a realistic value of $m_u/m_d \simeq 0.5$ one finds CSB at the 15\% level in the ratio $F^u_{K^+}(Q^2)/F^d_{K^0}(Q^2)$ at $Q^2 \simeq 10\,$GeV$^2$. The analogous changes in the quark distribution functions are considerably smaller in magnitude, reaching 3\% as $x\rightarrow 1$ in the pion ratio $u_{\pi^+}(x)/\bar{d}_{\pi^+}(x)$, compared with just 1\% in the ratio $u_{K^+}(x)/d_{K^0}(x)$ for the kaon. Testing these predictions presents considerable experimental challenges. Perhaps the most promising was outlined in Ref.~\cite{Londergan:1994gr} some time ago. By constructing the difference between four times the Drell-Yan cross section for $\pi^{+}$ and the corresponding cross section for $\pi^{-}$ on the deuteron ($4\sigma_{\pi^{+} D}^{DY} - \sigma_{\pi^{-} D}^{DY}$) and dividing by the average of these two terms ($[4\sigma_{\pi^{+} D}^{DY} + \sigma_{\pi^{-} D}^{DY}]/2$), one finds (at leading order and in the valence regions for both particles) a sum of just two terms. The first involves only the CSB distributions in the nucleon, evaluated at the Bjorken variable for the interacting parton in the deuteron ($x_1$). The second involves only the CSB quantity $\delta q_{\pi^+}(x) = \bar{d}_{\pi^+}(x) - u_{\pi^+}(x)$, divided by $\bar{d}_{\pi^+}(x)$ and evaluated at the Bjorken variable for the interacting parton in the pion ($x_2$). The complete separation of the nucleon and pion CSB terms makes this an attractive possibility. It would also be of interest to explore the expected degree of CSB in these systems using other realistic models and lattice QCD.

\vspace*{1em}
\begin{acknowledgments}
This work was supported by the U.S. Department of Energy, Office of Science, Office of Nuclear Physics, Contract No. DE-AC02-06CH11357; Laboratory Directed Research and Development (LDRD) funding from Argonne National Laboratory, Projects No. 2016-098-N0 and No. 2017-058-N0; and the University of Adelaide and the Australian Research Council through the ARC Centre of Excellence for Particle Physics at the Terascale (CE110001104) and Discovery Project DP151103101. P.~T.~P.~H was supported by the Young Scientist Training program of APCTP.
\end{acknowledgments}


\begin{thebibliography}{59}%
\makeatletter
\providecommand \@ifxundefined [1]{%
 \@ifx{#1\undefined}
}%
\providecommand \@ifnum [1]{%
 \ifnum #1\expandafter \@firstoftwo
 \else \expandafter \@secondoftwo
 \fi
}%
\providecommand \@ifx [1]{%
 \ifx #1\expandafter \@firstoftwo
 \else \expandafter \@secondoftwo
 \fi
}%
\providecommand \natexlab [1]{#1}%
\providecommand \enquote  [1]{``#1''}%
\providecommand \bibnamefont  [1]{#1}%
\providecommand \bibfnamefont [1]{#1}%
\providecommand \citenamefont [1]{#1}%
\providecommand \href@noop [0]{\@secondoftwo}%
\providecommand \href [0]{\begingroup \@sanitize@url \@href}%
\providecommand \@href[1]{\@@startlink{#1}\@@href}%
\providecommand \@@href[1]{\endgroup#1\@@endlink}%
\providecommand \@sanitize@url [0]{\catcode `\\12\catcode `\$12\catcode
  `\&12\catcode `\#12\catcode `\^12\catcode `\_12\catcode `\%12\relax}%
\providecommand \@@startlink[1]{}%
\providecommand \@@endlink[0]{}%
\providecommand \url  [0]{\begingroup\@sanitize@url \@url }%
\providecommand \@url [1]{\endgroup\@href {#1}{\urlprefix }}%
\providecommand \urlprefix  [0]{URL }%
\providecommand \Eprint [0]{\href }%
\providecommand \doibase [0]{http://dx.doi.org/}%
\providecommand \selectlanguage [0]{\@gobble}%
\providecommand \bibinfo  [0]{\@secondoftwo}%
\providecommand \bibfield  [0]{\@secondoftwo}%
\providecommand \translation [1]{[#1]}%
\providecommand \BibitemOpen [0]{}%
\providecommand \bibitemStop [0]{}%
\providecommand \bibitemNoStop [0]{.\EOS\space}%
\providecommand \EOS [0]{\spacefactor3000\relax}%
\providecommand \BibitemShut  [1]{\csname bibitem#1\endcsname}%
\let\auto@bib@innerbib\@empty
\bibitem [{\citenamefont {Miller}\ \emph {et~al.}(1990)\citenamefont {Miller},
  \citenamefont {Nefkens},\ and\ \citenamefont {Slaus}}]{Miller:1990iz}%
  \BibitemOpen
  \bibfield  {author} {\bibinfo {author} {\bibfnamefont {G.~A.}\ \bibnamefont
  {Miller}}, \bibinfo {author} {\bibfnamefont {B.~M.~K.}\ \bibnamefont
  {Nefkens}}\ and\ \bibinfo {author} {\bibfnamefont {I.}~\bibnamefont
  {Slaus}},\ }\href {\doibase 10.1016/0370-1573(90)90102-8} {\bibfield
  {journal} {\bibinfo  {journal} {Phys. Rept.}\ }\textbf {\bibinfo {volume}
  {194}},\ \bibinfo {pages} {1} (\bibinfo {year} {1990})}\BibitemShut {NoStop}%
\bibitem [{\citenamefont {Slaus}\ \emph {et~al.}(1991)\citenamefont {Slaus},
  \citenamefont {Nefkens},\ and\ \citenamefont {Miller}}]{Slaus:1990nn}%
  \BibitemOpen
  \bibfield  {author} {\bibinfo {author} {\bibfnamefont {I.}~\bibnamefont
  {Slaus}}, \bibinfo {author} {\bibfnamefont {B.~M.~K.}\ \bibnamefont
  {Nefkens}}\ and\ \bibinfo {author} {\bibfnamefont {G.~A.}\ \bibnamefont
  {Miller}},\ }\href {\doibase 10.1016/0168-583X(91)96077-X} {\bibfield
  {journal} {\bibinfo  {journal} {Nucl. Instrum. Meth.}\ }\textbf {\bibinfo
  {volume} {B56-57}},\ \bibinfo {pages} {489} (\bibinfo {year}
  {1991})}\BibitemShut {NoStop}%
\bibitem [{\citenamefont {Miller}\ \emph {et~al.}(2006)\citenamefont {Miller},
  \citenamefont {Opper},\ and\ \citenamefont {Stephenson}}]{Miller:2006tv}%
  \BibitemOpen
  \bibfield  {author} {\bibinfo {author} {\bibfnamefont {G.~A.}\ \bibnamefont
  {Miller}}, \bibinfo {author} {\bibfnamefont {A.~K.}\ \bibnamefont {Opper}}\
  and\ \bibinfo {author} {\bibfnamefont {E.~J.}\ \bibnamefont {Stephenson}},\
  }\href {\doibase 10.1146/annurev.nucl.56.080805.140446} {\bibfield  {journal}
  {\bibinfo  {journal} {Ann. Rev. Nucl. Part. Sci.}\ }\textbf {\bibinfo
  {volume} {56}},\ \bibinfo {pages} {253} (\bibinfo {year} {2006})}\Eprint
  {http://arxiv.org/abs/nucl-ex/0602021} {~[arXiv:nucl-ex/0602021
  [nucl-ex]]}\BibitemShut {NoStop}%
\bibitem [{\citenamefont {Londergan}\ \emph {et~al.}(2010)\citenamefont
  {Londergan}, \citenamefont {Peng},\ and\ \citenamefont
  {Thomas}}]{Londergan:2009kj}%
  \BibitemOpen
  \bibfield  {author} {\bibinfo {author} {\bibfnamefont {J.~T.}\ \bibnamefont
  {Londergan}}, \bibinfo {author} {\bibfnamefont {J.~C.}\ \bibnamefont {Peng}}\
  and\ \bibinfo {author} {\bibfnamefont {A.~W.}\ \bibnamefont {Thomas}},\
  }\href {\doibase 10.1103/RevModPhys.82.2009} {\bibfield  {journal} {\bibinfo
  {journal} {Rev. Mod. Phys.}\ }\textbf {\bibinfo {volume} {82}},\ \bibinfo
  {pages} {2009} (\bibinfo {year} {2010})}\Eprint
  {http://arxiv.org/abs/0907.2352} {~[arXiv:0907.2352 [hep-ph]]}\BibitemShut
  {NoStop}%
\bibitem [{\citenamefont {Gasser}\ and\ \citenamefont
  {Leutwyler}(1982)}]{Gasser:1982ap}%
  \BibitemOpen
  \bibfield  {author} {\bibinfo {author} {\bibfnamefont {J.}~\bibnamefont
  {Gasser}}\ and\ \bibinfo {author} {\bibfnamefont {H.}~\bibnamefont
  {Leutwyler}},\ }\href {\doibase 10.1016/0370-1573(82)90035-7} {\bibfield
  {journal} {\bibinfo  {journal} {Phys. Rept.}\ }\textbf {\bibinfo {volume}
  {87}},\ \bibinfo {pages} {77} (\bibinfo {year} {1982})}\BibitemShut {NoStop}%
\bibitem [{\citenamefont {Borsanyi}\ \emph {et~al.}(2013)\citenamefont
  {Borsanyi} \emph {et~al.}}]{Borsanyi:2013lga}%
  \BibitemOpen
  \bibfield  {author} {\bibinfo {author} {\bibfnamefont {S.}~\bibnamefont
  {Borsanyi}} \emph {et~al.} (\bibinfo {collaboration}
  {Budapest-Marseille-Wuppertal Collaboration}),\ }\href {\doibase
  10.1103/PhysRevLett.111.252001} {\bibfield  {journal} {\bibinfo  {journal}
  {Phys. Rev. Lett.}\ }\textbf {\bibinfo {volume} {111}},\ \bibinfo {pages}
  {252001} (\bibinfo {year} {2013})}\Eprint {http://arxiv.org/abs/1306.2287}
  {~[arXiv:1306.2287 [hep-lat]]}\BibitemShut {NoStop}%
\bibitem [{\citenamefont {Horsley}\ \emph
  {et~al.}(2016{\natexlab{a}})\citenamefont {Horsley} \emph
  {et~al.}}]{Horsley:2015vla}%
  \BibitemOpen
  \bibfield  {author} {\bibinfo {author} {\bibfnamefont {R.}~\bibnamefont
  {Horsley}} \emph {et~al.},\ }\href {\doibase 10.1007/JHEP04(2016)093}
  {\bibfield  {journal} {\bibinfo  {journal} {J. High Energy Phys.}\ }\textbf {\bibinfo
  {volume} {04}},\ \bibinfo {pages} {093} (\bibinfo {year}
  {2016}{\natexlab{a}})}\Eprint {http://arxiv.org/abs/1509.00799}
  {~[arXiv:1509.00799 [hep-lat]]}\BibitemShut {NoStop}%
\bibitem [{\citenamefont {Horsley}\ \emph
  {et~al.}(2016{\natexlab{b}})\citenamefont {Horsley} \emph
  {et~al.}}]{Horsley:2015eaa}%
  \BibitemOpen
  \bibfield  {author} {\bibinfo {author} {\bibfnamefont {R.}~\bibnamefont
  {Horsley}} \emph {et~al.},\ }\href {\doibase 10.1088/0954-3899/43/10/10LT02}
  {\bibfield  {journal} {\bibinfo  {journal} {J. Phys.}\ }\textbf {\bibinfo
  {volume} {G43}},\ \bibinfo {pages} {10LT02} (\bibinfo {year}
  {2016}{\natexlab{b}})}\Eprint {http://arxiv.org/abs/1508.06401}
  {~[arXiv:1508.06401 [hep-lat]]}\BibitemShut {NoStop}%
\bibitem [{\citenamefont {Kubis}(2011)}]{Kubis:2009cp}%
  \BibitemOpen
  \bibfield  {author} {\bibinfo {author} {\bibfnamefont {B.}~\bibnamefont
  {Kubis}},\ }\bibfield  {booktitle} {\emph {\bibinfo {booktitle} {{in 4th
  International Workshop on From Parity Violation to Hadronic Structure and
  More (PAVI09) Bar Harbor, Maine, June 22-26, 2009}}},\ }\href {\doibase
  10.1007/s10751-011-0284-x} {\bibfield  {journal} {\bibinfo  {journal}
  {Hyperfine Interact.}\ }\textbf {\bibinfo {volume} {201}},\ \bibinfo {pages}
  {7} (\bibinfo {year} {2011})}\Eprint {http://arxiv.org/abs/0910.2800}
  {~[arXiv:0910.2800 [nucl-th]]}\BibitemShut {NoStop}%
\bibitem [{\citenamefont {Shanahan}\ \emph {et~al.}(2015)\citenamefont
  {Shanahan}, \citenamefont {Horsley}, \citenamefont {Nakamura}, \citenamefont
  {Pleiter}, \citenamefont {Rakow}, \citenamefont {Schierholz}, \citenamefont
  {Stüben}, \citenamefont {Thomas}, \citenamefont {Young},\ and\ \citenamefont
  {Zanotti}}]{Shanahan:2015caa}%
  \BibitemOpen
  \bibfield  {author} {\bibinfo {author} {\bibfnamefont {P.~E.}\ \bibnamefont
  {Shanahan}}, \bibinfo {author} {\bibfnamefont {R.}~\bibnamefont {Horsley}},
  \bibinfo {author} {\bibfnamefont {Y.}~\bibnamefont {Nakamura}}, \bibinfo
  {author} {\bibfnamefont {D.}~\bibnamefont {Pleiter}}, \bibinfo {author}
  {\bibfnamefont {P.~E.~L.}\ \bibnamefont {Rakow}}, \bibinfo {author}
  {\bibfnamefont {G.}~\bibnamefont {Schierholz}}, \bibinfo {author}
  {\bibfnamefont {H.}~\bibnamefont {Stüben}}, \bibinfo {author} {\bibfnamefont
  {A.~W.}\ \bibnamefont {Thomas}}, \bibinfo {author} {\bibfnamefont {R.~D.}\
  \bibnamefont {Young}}\ and\ \bibinfo {author} {\bibfnamefont {J.~M.}\
  \bibnamefont {Zanotti}},\ }\href {\doibase 10.1103/PhysRevD.91.113006}
  {\bibfield  {journal} {\bibinfo  {journal} {Phys. Rev.}\ }\textbf {\bibinfo
  {volume} {D91}},\ \bibinfo {pages} {113006} (\bibinfo {year} {2015})}\Eprint
  {http://arxiv.org/abs/1503.01142} {~[arXiv:1503.01142 [hep-lat]]}\BibitemShut
  {NoStop}%
\bibitem [{\citenamefont {Zeller}\ \emph {et~al.}(2002)\citenamefont {Zeller}
  \emph {et~al.}}]{Zeller:2001hh}%
  \BibitemOpen
  \bibfield  {author} {\bibinfo {author} {\bibfnamefont {G.~P.}\ \bibnamefont
  {Zeller}} \emph {et~al.} (\bibinfo {collaboration} {NuTeV Collaboration}),\ }\href
  {\doibase 10.1103/PhysRevLett.88.091802} {\bibfield  {journal} {\bibinfo
  {journal} {Phys. Rev. Lett.}\ }\textbf {\bibinfo {volume} {88}},\ \bibinfo
  {pages} {091802} (\bibinfo {year} {2002})}\bibinfo {note} {[Erratum: Phys.
  Rev. Lett.90,239902(2003)]},\ \Eprint {http://arxiv.org/abs/hep-ex/0110059}
  {~[arXiv:hep-ex/0110059 [hep-ex]]}\BibitemShut {NoStop}%
\bibitem [{\citenamefont {Cloët}\ \emph {et~al.}(2009)\citenamefont {Cloët},
  \citenamefont {Bentz},\ and\ \citenamefont {Thomas}}]{Cloet:2009qs}%
  \BibitemOpen
  \bibfield  {author} {\bibinfo {author} {\bibfnamefont {I.~C.}\ \bibnamefont
  {Cloët}}, \bibinfo {author} {\bibfnamefont {W.}~\bibnamefont {Bentz}}\ and\
  \bibinfo {author} {\bibfnamefont {A.~W.}\ \bibnamefont {Thomas}},\ }\href
  {\doibase 10.1103/PhysRevLett.102.252301} {\bibfield  {journal} {\bibinfo
  {journal} {Phys. Rev. Lett.}\ }\textbf {\bibinfo {volume} {102}},\ \bibinfo
  {pages} {252301} (\bibinfo {year} {2009})}\Eprint
  {http://arxiv.org/abs/0901.3559} {~[arXiv:0901.3559 [nucl-th]]}\BibitemShut
  {NoStop}%
\bibitem [{\citenamefont {Martin}\ \emph {et~al.}(2004)\citenamefont {Martin},
  \citenamefont {Roberts}, \citenamefont {Stirling},\ and\ \citenamefont
  {Thorne}}]{Martin:2003sk}%
  \BibitemOpen
  \bibfield  {author} {\bibinfo {author} {\bibfnamefont {A.~D.}\ \bibnamefont
  {Martin}}, \bibinfo {author} {\bibfnamefont {R.~G.}\ \bibnamefont {Roberts}},
  \bibinfo {author} {\bibfnamefont {W.~J.}\ \bibnamefont {Stirling}}\ and\
  \bibinfo {author} {\bibfnamefont {R.~S.}\ \bibnamefont {Thorne}},\ }\href
  {\doibase 10.1140/epjc/s2004-01825-2} {\bibfield  {journal} {\bibinfo
  {journal} {Eur. Phys. J.}\ }\textbf {\bibinfo {volume} {C35}},\ \bibinfo
  {pages} {325} (\bibinfo {year} {2004})}\Eprint
  {http://arxiv.org/abs/hep-ph/0308087} {~[arXiv:hep-ph/0308087
  [hep-ph]]}\BibitemShut {NoStop}%
\bibitem [{\citenamefont {Londergan}\ \emph {et~al.}(2005)\citenamefont
  {Londergan}, \citenamefont {Murdock},\ and\ \citenamefont
  {Thomas}}]{Londergan:2005ht}%
  \BibitemOpen
  \bibfield  {author} {\bibinfo {author} {\bibfnamefont {J.~T.}\ \bibnamefont
  {Londergan}}, \bibinfo {author} {\bibfnamefont {D.~P.}\ \bibnamefont
  {Murdock}}\ and\ \bibinfo {author} {\bibfnamefont {A.~W.}\ \bibnamefont
  {Thomas}},\ }\href {\doibase 10.1103/PhysRevD.72.036010} {\bibfield
  {journal} {\bibinfo  {journal} {Phys. Rev.}\ }\textbf {\bibinfo {volume}
  {D72}},\ \bibinfo {pages} {036010} (\bibinfo {year} {2005})}\Eprint
  {http://arxiv.org/abs/hep-ph/0507029} {~[arXiv:hep-ph/0507029
  [hep-ph]]}\BibitemShut {NoStop}%
\bibitem [{\citenamefont {Bentz}\ \emph {et~al.}(2010)\citenamefont {Bentz},
  \citenamefont {Cloët}, \citenamefont {Londergan},\ and\ \citenamefont
  {Thomas}}]{Bentz:2009yy}%
  \BibitemOpen
  \bibfield  {author} {\bibinfo {author} {\bibfnamefont {W.}~\bibnamefont
  {Bentz}}, \bibinfo {author} {\bibfnamefont {I.~C.}\ \bibnamefont {Cloët}},
  \bibinfo {author} {\bibfnamefont {J.~T.}\ \bibnamefont {Londergan}}\ and\
  \bibinfo {author} {\bibfnamefont {A.~W.}\ \bibnamefont {Thomas}},\ }\href
  {\doibase 10.1016/j.physletb.2010.09.001} {\bibfield  {journal} {\bibinfo
  {journal} {Phys. Lett.}\ }\textbf {\bibinfo {volume} {B693}},\ \bibinfo
  {pages} {462} (\bibinfo {year} {2010})}\Eprint
  {http://arxiv.org/abs/0908.3198} {~[arXiv:0908.3198 [nucl-th]]}\BibitemShut
  {NoStop}%
\bibitem [{\citenamefont {Wang}\ \emph {et~al.}(2016)\citenamefont {Wang},
  \citenamefont {Thomas},\ and\ \citenamefont {Young}}]{Wang:2015msk}%
  \BibitemOpen
  \bibfield  {author} {\bibinfo {author} {\bibfnamefont {X.~G.}\ \bibnamefont
  {Wang}}, \bibinfo {author} {\bibfnamefont {A.~W.}\ \bibnamefont {Thomas}}\
  and\ \bibinfo {author} {\bibfnamefont {R.~D.}\ \bibnamefont {Young}},\ }\href
  {\doibase 10.1016/j.physletb.2015.12.062} {\bibfield  {journal} {\bibinfo
  {journal} {Phys. Lett.}\ }\textbf {\bibinfo {volume} {B753}},\ \bibinfo
  {pages} {595} (\bibinfo {year} {2016})}\Eprint
  {http://arxiv.org/abs/1512.04139} {~[arXiv:1512.04139 [nucl-th]]}\BibitemShut
  {NoStop}%
\bibitem [{\citenamefont {Shanahan}\ \emph {et~al.}(2013)\citenamefont
  {Shanahan}, \citenamefont {Thomas},\ and\ \citenamefont
  {Young}}]{Shanahan:2013vla}%
  \BibitemOpen
  \bibfield  {author} {\bibinfo {author} {\bibfnamefont {P.~E.}\ \bibnamefont
  {Shanahan}}, \bibinfo {author} {\bibfnamefont {A.~W.}\ \bibnamefont
  {Thomas}}\ and\ \bibinfo {author} {\bibfnamefont {R.~D.}\ \bibnamefont
  {Young}},\ }\href {\doibase 10.1103/PhysRevD.87.094515} {\bibfield  {journal}
  {\bibinfo  {journal} {Phys. Rev.}\ }\textbf {\bibinfo {volume} {D87}},\
  \bibinfo {pages} {094515} (\bibinfo {year} {2013})}\Eprint
  {http://arxiv.org/abs/1303.4806} {~[arXiv:1303.4806 [nucl-th]]}\BibitemShut
  {NoStop}%
\bibitem [{\citenamefont {Rodionov}\ \emph {et~al.}(1994)\citenamefont
  {Rodionov}, \citenamefont {Thomas},\ and\ \citenamefont
  {Londergan}}]{Rodionov:1994cg}%
  \BibitemOpen
  \bibfield  {author} {\bibinfo {author} {\bibfnamefont {E.~N.}\ \bibnamefont
  {Rodionov}}, \bibinfo {author} {\bibfnamefont {A.~W.}\ \bibnamefont
  {Thomas}}\ and\ \bibinfo {author} {\bibfnamefont {J.~T.}\ \bibnamefont
  {Londergan}},\ }\href {\doibase 10.1142/S0217732394001659} {\bibfield
  {journal} {\bibinfo  {journal} {Mod. Phys. Lett.}\ }\textbf {\bibinfo
  {volume} {A9}},\ \bibinfo {pages} {1799} (\bibinfo {year}
  {1994})}\BibitemShut {NoStop}%
\bibitem [{\citenamefont {Sather}(1992)}]{Sather:1991je}%
  \BibitemOpen
  \bibfield  {author} {\bibinfo {author} {\bibfnamefont {E.}~\bibnamefont
  {Sather}},\ }\href {\doibase 10.1016/0370-2693(92)92011-5} {\bibfield
  {journal} {\bibinfo  {journal} {Phys. Lett.}\ }\textbf {\bibinfo {volume}
  {B274}},\ \bibinfo {pages} {433} (\bibinfo {year} {1992})}\BibitemShut
  {NoStop}%
\bibitem [{\citenamefont {Londergan}\ and\ \citenamefont
  {Thomas}(1998)}]{Londergan:1998ai}%
  \BibitemOpen
  \bibfield  {author} {\bibinfo {author} {\bibfnamefont {J.~T.}\ \bibnamefont
  {Londergan}}\ and\ \bibinfo {author} {\bibfnamefont {A.~W.}\ \bibnamefont
  {Thomas}},\ }\href {\doibase 10.1016/S0146-6410(98)00055-6} {\bibfield
  {journal} {\bibinfo  {journal} {Prog. Part. Nucl. Phys.}\ }\textbf {\bibinfo
  {volume} {41}},\ \bibinfo {pages} {49} (\bibinfo {year} {1998})}\Eprint
  {http://arxiv.org/abs/hep-ph/9806510} {~[arXiv:hep-ph/9806510
  [hep-ph]]}\BibitemShut {NoStop}%
\bibitem [{\citenamefont {Nolen}\ and\ \citenamefont
  {Schiffer}(1969)}]{Nolen:1969ms}%
  \BibitemOpen
  \bibfield  {author} {\bibinfo {author} {\bibfnamefont {J.~A.}\ \bibnamefont
  {Nolen}, \bibfnamefont {Jr.}}\ and\ \bibinfo {author} {\bibfnamefont {J.~P.}\
  \bibnamefont {Schiffer}},\ }\href {\doibase
  10.1146/annurev.ns.19.120169.002351} {\bibfield  {journal} {\bibinfo
  {journal} {Ann. Rev. Nucl. Part. Sci.}\ }\textbf {\bibinfo {volume} {19}},\
  \bibinfo {pages} {471} (\bibinfo {year} {1969})}\BibitemShut {NoStop}%
\bibitem [{\citenamefont {Henley}\ and\ \citenamefont
  {Krein} (1989)}]{Henley:1989vi}%
  \BibitemOpen
  \bibfield  {author} {\bibinfo {author} {\bibfnamefont {E.~M.}~\bibnamefont
  {Henley}} \ and\ \bibinfo {author} {\bibfnamefont {G.}~\bibnamefont {Krein}},\ }\href
  {\doibase 10.1103/PhysRevLett.62.2586} {\bibfield  {journal} {\bibinfo
  {journal} {Phys. Rev. Lett.}\ }\textbf {\bibinfo {volume} {62}},\ \bibinfo {pages}
  {2586} (\bibinfo {year} {1989})}\BibitemShut {NoStop}%
\bibitem [{\citenamefont {Hatsuda}\ \emph {et~al.}(1990)\citenamefont
  {Hatsuda}, \citenamefont {Hogaasen},\ and\ \citenamefont
  {Prakash}}]{Hatsuda:1990pj}%
  \BibitemOpen
  \bibfield  {author} {\bibinfo {author} {\bibfnamefont {T.}~\bibnamefont
  {Hatsuda}}, \bibinfo {author} {\bibfnamefont {H.}~\bibnamefont {Hogaasen}}\
  and\ \bibinfo {author} {\bibfnamefont {M.}~\bibnamefont {Prakash}},\ }\href
  {\doibase 10.1103/PhysRevC.42.2212} {\bibfield  {journal} {\bibinfo
  {journal} {Phys. Rev.}\ }\textbf {\bibinfo {volume} {C42}},\ \bibinfo {pages}
  {2212} (\bibinfo {year} {1990})}\BibitemShut {NoStop}%
\bibitem [{\citenamefont {Saito}\ and\ \citenamefont
  {Thomas}(1994)}]{Saito:1994tq}%
  \BibitemOpen
  \bibfield  {author} {\bibinfo {author} {\bibfnamefont {K.}~\bibnamefont
  {Saito}}\ and\ \bibinfo {author} {\bibfnamefont {A.~W.}\ \bibnamefont
  {Thomas}},\ }\href {\doibase 10.1016/0370-2693(94)91551-2} {\bibfield
  {journal} {\bibinfo  {journal} {Phys. Lett.}\ }\textbf {\bibinfo {volume}
  {B335}},\ \bibinfo {pages} {17} (\bibinfo {year} {1994})}\Eprint
  {http://arxiv.org/abs/nucl-th/9405009} {~[arXiv:nucl-th/9405009
  [nucl-th]]}\BibitemShut {NoStop}%
\bibitem [{\citenamefont {Miller}(2013)}]{Miller:2013nea}%
  \BibitemOpen
  \bibfield  {author} {\bibinfo {author} {\bibfnamefont {G.~A.}\ \bibnamefont
  {Miller}},\ }\bibfield  {booktitle} {\emph {\bibinfo {booktitle}
  {{in Workshop to Explore Physics Opportunities with Intense,
  Polarized Electron Beams at 50-300 MeV: Cambridge, MA, March 14-16,
  2013, edited by R. Milner, R. Carlini, and F. Maas,}}},\ }\href {\doibase 10.1063/1.4829385} {\bibfield  {journal} {\bibinfo
   {journal} {AIP Conf. Proc.}\ }\textbf {\bibinfo {volume} {1563}},\ \bibinfo
  {pages} {102} (\bibinfo {year} {2013})}\Eprint
  {http://arxiv.org/abs/1309.0879} {~[arXiv:1309.0879 [nucl-th]]}\BibitemShut
  {NoStop}%
\bibitem [{\citenamefont {Londergan}\ \emph {et~al.}(1996)\citenamefont
  {Londergan}, \citenamefont {Pang},\ and\ \citenamefont
  {Thomas}}]{Londergan:1996vf}%
  \BibitemOpen
  \bibfield  {author} {\bibinfo {author} {\bibfnamefont {J.~T.}\ \bibnamefont
  {Londergan}}, \bibinfo {author} {\bibfnamefont {A.}~\bibnamefont {Pang}}\
  and\ \bibinfo {author} {\bibfnamefont {A.~W.}\ \bibnamefont {Thomas}},\
  }\href {\doibase 10.1103/PhysRevD.54.3154} {\bibfield  {journal} {\bibinfo
  {journal} {Phys. Rev.}\ }\textbf {\bibinfo {volume} {D54}},\ \bibinfo {pages}
  {3154} (\bibinfo {year} {1996})}\Eprint {http://arxiv.org/abs/hep-ph/9604446}
  {~[arXiv:hep-ph/9604446 [hep-ph]]}\BibitemShut {NoStop}%
\bibitem [{\citenamefont {Accardi}\ \emph {et~al.}(2016)\citenamefont {Accardi}
  \emph {et~al.}}]{Accardi:2012qut}%
  \BibitemOpen
  \bibfield  {author} {\bibinfo {author} {\bibfnamefont {A.}~\bibnamefont
  {Accardi}} \emph {et~al.},\ }\href {\doibase 10.1140/epja/i2016-16268-9}
  {\bibfield  {journal} {\bibinfo  {journal} {Eur. Phys. J.}\ }\textbf
  {\bibinfo {volume} {A52}},\ \bibinfo {pages} {268} (\bibinfo {year}
  {2016})}\Eprint {http://arxiv.org/abs/1212.1701} {~[arXiv:1212.1701
  [nucl-ex]]}\BibitemShut {NoStop}%
\bibitem [{\citenamefont {Boer}\ \emph {et~al.}(2011)\citenamefont {Boer} \emph
  {et~al.}}]{Boer:2011fh}%
  \BibitemOpen
  \bibfield  {author} {\bibinfo {author} {\bibfnamefont {D.}~\bibnamefont
  {Boer}} \emph {et~al.},\ }\href@noop {} {\ }\Eprint
  {http://arxiv.org/abs/1108.1713} {arXiv:1108.1713 [nucl-th]}\BibitemShut
  {NoStop}%
\bibitem [{\citenamefont {Londergan}\ \emph {et~al.}(1994)\citenamefont
  {Londergan}, \citenamefont {Carvey}, \citenamefont {Liu}, \citenamefont
  {Rodionov},\ and\ \citenamefont {Thomas}}]{Londergan:1994gr}%
  \BibitemOpen
  \bibfield  {author} {\bibinfo {author} {\bibfnamefont {J.~T.}\ \bibnamefont
  {Londergan}}, \bibinfo {author} {\bibfnamefont {G.~T.}\ \bibnamefont
  {Carvey}}, \bibinfo {author} {\bibfnamefont {G.~Q.}\ \bibnamefont {Liu}},
  \bibinfo {author} {\bibfnamefont {E.~N.}\ \bibnamefont {Rodionov}}\ and\
  \bibinfo {author} {\bibfnamefont {A.~W.}\ \bibnamefont {Thomas}},\ }\href
  {\doibase 10.1016/0370-2693(94)91306-4} {\bibfield  {journal} {\bibinfo
  {journal} {Phys. Lett.}\ }\textbf {\bibinfo {volume} {B340}},\ \bibinfo
  {pages} {115} (\bibinfo {year} {1994})}\Eprint
  {http://arxiv.org/abs/hep-ph/9503221} {~[arXiv:hep-ph/9503221
  [hep-ph]]}\BibitemShut {NoStop}%
\bibitem [{\citenamefont {Conway}\ \emph {et~al.}(1989)\citenamefont {Conway}
  \emph {et~al.}}]{Conway:1989fs}%
  \BibitemOpen
  \bibfield  {author} {\bibinfo {author} {\bibfnamefont {J.~S.}\ \bibnamefont
  {Conway}} \emph {et~al.},\ }\href {\doibase 10.1103/PhysRevD.39.92}
  {\bibfield  {journal} {\bibinfo  {journal} {Phys. Rev.}\ }\textbf {\bibinfo
  {volume} {D39}},\ \bibinfo {pages} {92} (\bibinfo {year} {1989})}\BibitemShut
  {NoStop}%
\bibitem [{\citenamefont {Nambu}\ and\ \citenamefont
  {Jona-Lasinio}(1961{\natexlab{a}})}]{Nambu:1961fr}%
  \BibitemOpen
  \bibfield  {author} {\bibinfo {author} {\bibfnamefont {Y.}~\bibnamefont
  {Nambu}}\ and\ \bibinfo {author} {\bibfnamefont {G.}~\bibnamefont
  {Jona-Lasinio}},\ }\href {\doibase 10.1103/PhysRev.124.246} {\bibfield
  {journal} {\bibinfo  {journal} {Phys. Rev.}\ }\textbf {\bibinfo {volume}
  {124}},\ \bibinfo {pages} {246} (\bibinfo {year}
  {1961}{\natexlab{a}})}\BibitemShut {NoStop}%
\bibitem [{\citenamefont {Nambu}\ and\ \citenamefont
  {Jona-Lasinio}(1961{\natexlab{b}})}]{Nambu:1961tp}%
  \BibitemOpen
  \bibfield  {author} {\bibinfo {author} {\bibfnamefont {Y.}~\bibnamefont
  {Nambu}}\ and\ \bibinfo {author} {\bibfnamefont {G.}~\bibnamefont
  {Jona-Lasinio}},\ }\href {\doibase 10.1103/PhysRev.122.345} {\bibfield
  {journal} {\bibinfo  {journal} {Phys. Rev.}\ }\textbf {\bibinfo {volume}
  {122}},\ \bibinfo {pages} {345} (\bibinfo {year}
  {1961}{\natexlab{b}})}\BibitemShut {NoStop}%
\bibitem [{\citenamefont {Klevansky}(1992)}]{Klevansky:1992qe}%
  \BibitemOpen
  \bibfield  {author} {\bibinfo {author} {\bibfnamefont {S.}~\bibnamefont
  {Klevansky}},\ }\href {\doibase 10.1103/RevModPhys.64.649} {\bibfield
  {journal} {\bibinfo  {journal} {Rev. Mod. Phys.}\ }\textbf {\bibinfo {volume}
  {64}},\ \bibinfo {pages} {649} (\bibinfo {year} {1992})}\BibitemShut
  {NoStop}%
\bibitem [{\citenamefont {Vogl}\ and\ \citenamefont
  {Weise}(1991)}]{Vogl:1991qt}%
  \BibitemOpen
  \bibfield  {author} {\bibinfo {author} {\bibfnamefont {U.}~\bibnamefont
  {Vogl}}\ and\ \bibinfo {author} {\bibfnamefont {W.}~\bibnamefont {Weise}},\
  }\href {\doibase 10.1016/0146-6410(91)90005-9} {\bibfield  {journal}
  {\bibinfo  {journal} {Prog. Part. Nucl. Phys.}\ }\textbf {\bibinfo {volume}
  {27}},\ \bibinfo {pages} {195} (\bibinfo {year} {1991})}\BibitemShut
  {NoStop}%
\bibitem [{\citenamefont {Cloët}\ \emph {et~al.}(2014)\citenamefont {Cloët},
  \citenamefont {Bentz},\ and\ \citenamefont {Thomas}}]{Cloet:2014rja}%
  \BibitemOpen
  \bibfield  {author} {\bibinfo {author} {\bibfnamefont {I.~C.}\ \bibnamefont
  {Cloët}}, \bibinfo {author} {\bibfnamefont {W.}~\bibnamefont {Bentz}}\ and\
  \bibinfo {author} {\bibfnamefont {A.~W.}\ \bibnamefont {Thomas}},\ }\href
  {\doibase 10.1103/PhysRevC.90.045202} {\bibfield  {journal} {\bibinfo
  {journal} {Phys. Rev.}\ }\textbf {\bibinfo {volume} {C90}},\ \bibinfo {pages}
  {045202} (\bibinfo {year} {2014})}\Eprint {http://arxiv.org/abs/1405.5542}
  {~[arXiv:1405.5542 [nucl-th]]}\BibitemShut {NoStop}%
\bibitem [{\citenamefont {Schwinger}(1951)}]{Schwinger:1951nm}%
  \BibitemOpen
  \bibfield  {author} {\bibinfo {author} {\bibfnamefont {J.~S.}\ \bibnamefont
  {Schwinger}},\ }\href {\doibase 10.1103/PhysRev.82.664} {\bibfield  {journal}
  {\bibinfo  {journal} {Phys. Rev.}\ }\textbf {\bibinfo {volume} {82}},\
  \bibinfo {pages} {664} (\bibinfo {year} {1951})}\BibitemShut {NoStop}%
\bibitem [{\citenamefont {Ebert}\ \emph {et~al.}(1996)\citenamefont {Ebert},
  \citenamefont {Feldmann},\ and\ \citenamefont {Reinhardt}}]{Ebert:1996vx}%
  \BibitemOpen
  \bibfield  {author} {\bibinfo {author} {\bibfnamefont {D.}~\bibnamefont
  {Ebert}}, \bibinfo {author} {\bibfnamefont {T.}~\bibnamefont {Feldmann}}\
  and\ \bibinfo {author} {\bibfnamefont {H.}~\bibnamefont {Reinhardt}},\ }\href
  {\doibase 10.1016/0370-2693(96)01158-6} {\bibfield  {journal} {\bibinfo
  {journal} {Phys. Lett. B}\ }\textbf {\bibinfo {volume} {388}},\ \bibinfo
  {pages} {154} (\bibinfo {year} {1996})}\Eprint {http://arxiv.org/abs/9608223}
  {~[arXiv:9608223 [hep-ph]]}\BibitemShut {NoStop}%
\bibitem [{\citenamefont {Hellstern}\ \emph {et~al.}(1997)\citenamefont
  {Hellstern}, \citenamefont {Alkofer},\ and\ \citenamefont
  {Reinhardt}}]{Hellstern:1997nv}%
  \BibitemOpen
  \bibfield  {author} {\bibinfo {author} {\bibfnamefont {G.}~\bibnamefont
  {Hellstern}}, \bibinfo {author} {\bibfnamefont {R.}~\bibnamefont {Alkofer}}\
  and\ \bibinfo {author} {\bibfnamefont {H.}~\bibnamefont {Reinhardt}},\ }\href
  {\doibase 10.1016/S0375-9474(97)00412-0} {\bibfield  {journal} {\bibinfo
  {journal} {Nucl. Phys. A}\ }\textbf {\bibinfo {volume} {625}},\ \bibinfo
  {pages} {697} (\bibinfo {year} {1997})}\Eprint {http://arxiv.org/abs/9706551}
  {~[arXiv:9706551 [hep-ph]]}\BibitemShut {NoStop}%
\bibitem [{\citenamefont {Bentz}\ and\ \citenamefont
  {Thomas}(2001)}]{Bentz:2001vc}%
  \BibitemOpen
  \bibfield  {author} {\bibinfo {author} {\bibfnamefont {W.}~\bibnamefont
  {Bentz}}\ and\ \bibinfo {author} {\bibfnamefont {A.~W.}\ \bibnamefont
  {Thomas}},\ }\href {\doibase 10.1016/S0375-9474(01)01119-8} {\bibfield
  {journal} {\bibinfo  {journal} {Nucl. Phys. A}\ }\textbf {\bibinfo {volume}
  {696}},\ \bibinfo {pages} {138} (\bibinfo {year} {2001})}\Eprint
  {http://arxiv.org/abs/0105022} {~[arXiv:0105022 [nucl-th]]}\BibitemShut
  {NoStop}%
\bibitem [{\citenamefont {Vogl}\ \emph {et~al.}(1990)\citenamefont {Vogl},
  \citenamefont {Lutz}, \citenamefont {Klimt},\ and\ \citenamefont
  {Weise}}]{Vogl:1989ea}%
  \BibitemOpen
  \bibfield  {author} {\bibinfo {author} {\bibfnamefont {U.}~\bibnamefont
  {Vogl}}, \bibinfo {author} {\bibfnamefont {M.~F.~M.}\ \bibnamefont {Lutz}},
  \bibinfo {author} {\bibfnamefont {S.}~\bibnamefont {Klimt}}\ and\ \bibinfo
  {author} {\bibfnamefont {W.}~\bibnamefont {Weise}},\ }\href {\doibase
  10.1016/0375-9474(90)90124-5} {\bibfield  {journal} {\bibinfo  {journal}
  {Nucl. Phys.}\ }\textbf {\bibinfo {volume} {A516}},\ \bibinfo {pages} {469}
  (\bibinfo {year} {1990})}\BibitemShut {NoStop}%
\bibitem [{\citenamefont {Klimt}\ \emph {et~al.}(1990)\citenamefont {Klimt},
  \citenamefont {Lutz}, \citenamefont {Vogl},\ and\ \citenamefont
  {Weise}}]{Klimt:1989pm}%
  \BibitemOpen
  \bibfield  {author} {\bibinfo {author} {\bibfnamefont {S.}~\bibnamefont
  {Klimt}}, \bibinfo {author} {\bibfnamefont {M.~F.~M.}\ \bibnamefont {Lutz}},
  \bibinfo {author} {\bibfnamefont {U.}~\bibnamefont {Vogl}}\ and\ \bibinfo
  {author} {\bibfnamefont {W.}~\bibnamefont {Weise}},\ }\href {\doibase
  10.1016/0375-9474(90)90123-4} {\bibfield  {journal} {\bibinfo  {journal}
  {Nucl. Phys.}\ }\textbf {\bibinfo {volume} {A516}},\ \bibinfo {pages} {429}
  (\bibinfo {year} {1990})}\BibitemShut {NoStop}%
\bibitem [{\citenamefont {Ishii}\ \emph {et~al.}(1995)\citenamefont {Ishii},
  \citenamefont {Bentz},\ and\ \citenamefont {Yazaki}}]{Ishii:1995bu}%
  \BibitemOpen
  \bibfield  {author} {\bibinfo {author} {\bibfnamefont {N.}~\bibnamefont
  {Ishii}}, \bibinfo {author} {\bibfnamefont {W.}~\bibnamefont {Bentz}}\ and\
  \bibinfo {author} {\bibfnamefont {K.}~\bibnamefont {Yazaki}},\ }\href
  {\doibase 10.1016/0375-9474(95)00032-V} {\bibfield  {journal} {\bibinfo
  {journal} {Nucl. Phys. A}\ }\textbf {\bibinfo {volume} {587}},\ \bibinfo
  {pages} {617} (\bibinfo {year} {1995})}\BibitemShut {NoStop}%
\bibitem [{\citenamefont {Cloët}\ \emph {et~al.}(2008)\citenamefont {Cloët},
  \citenamefont {Bentz},\ and\ \citenamefont {Thomas}}]{Cloet:2007em}%
  \BibitemOpen
  \bibfield  {author} {\bibinfo {author} {\bibfnamefont {I.~C.}\ \bibnamefont
  {Cloët}}, \bibinfo {author} {\bibfnamefont {W.}~\bibnamefont {Bentz}}\ and\
  \bibinfo {author} {\bibfnamefont {A.~W.}\ \bibnamefont {Thomas}},\ }\href
  {\doibase 10.1016/j.physletb.2007.09.071} {\bibfield  {journal} {\bibinfo
  {journal} {Phys. Lett.}\ }\textbf {\bibinfo {volume} {B659}},\ \bibinfo
  {pages} {214} (\bibinfo {year} {2008})}\Eprint
  {http://arxiv.org/abs/0708.3246} {~[arXiv:0708.3246 [hep-ph]]}\BibitemShut
  {NoStop}%
\bibitem [{\citenamefont {Carrillo-Serrano}\ \emph {et~al.}(2014)\citenamefont
  {Carrillo-Serrano}, \citenamefont {Cloët},\ and\ \citenamefont
  {Thomas}}]{Carrillo-Serrano:2014zta}%
  \BibitemOpen
  \bibfield  {author} {\bibinfo {author} {\bibfnamefont {M.~E.}\ \bibnamefont
  {Carrillo-Serrano}}, \bibinfo {author} {\bibfnamefont {I.~C.}\ \bibnamefont
  {Cloët}}\ and\ \bibinfo {author} {\bibfnamefont {A.~W.}\ \bibnamefont
  {Thomas}},\ }\href {\doibase 10.1103/PhysRevC.90.064316} {\bibfield
  {journal} {\bibinfo  {journal} {Phys. Rev.}\ }\textbf {\bibinfo {volume}
  {C90}},\ \bibinfo {pages} {064316} (\bibinfo {year} {2014})}\Eprint
  {http://arxiv.org/abs/1409.1653} {~[arXiv:1409.1653 [nucl-th]]}\BibitemShut
  {NoStop}%
\bibitem [{\citenamefont {Ito}\ \emph {et~al.}(2009)\citenamefont {Ito},
  \citenamefont {Bentz}, \citenamefont {Cloët}, \citenamefont {Thomas},\ and\
  \citenamefont {Yazaki}}]{Ito:2009zc}%
  \BibitemOpen
  \bibfield  {author} {\bibinfo {author} {\bibfnamefont {T.}~\bibnamefont
  {Ito}}, \bibinfo {author} {\bibfnamefont {W.}~\bibnamefont {Bentz}}, \bibinfo
  {author} {\bibfnamefont {I.~C.}\ \bibnamefont {Cloët}}, \bibinfo {author}
  {\bibfnamefont {A.~W.}\ \bibnamefont {Thomas}}\ and\ \bibinfo {author}
  {\bibfnamefont {K.}~\bibnamefont {Yazaki}},\ }\href {\doibase
  10.1103/PhysRevD.80.074008} {\bibfield  {journal} {\bibinfo  {journal} {Phys.
  Rev.}\ }\textbf {\bibinfo {volume} {D80}},\ \bibinfo {pages} {074008}
  (\bibinfo {year} {2009})}\Eprint {http://arxiv.org/abs/0906.5362}
  {~[arXiv:0906.5362 [nucl-th]]}\BibitemShut {NoStop}%
\bibitem [{\citenamefont {Matevosyan}\ \emph {et~al.}(2012)\citenamefont
  {Matevosyan}, \citenamefont {Bentz}, \citenamefont {Cloët},\ and\
  \citenamefont {Thomas}}]{Matevosyan:2011vj}%
  \BibitemOpen
  \bibfield  {author} {\bibinfo {author} {\bibfnamefont {H.~H.}\ \bibnamefont
  {Matevosyan}}, \bibinfo {author} {\bibfnamefont {W.}~\bibnamefont {Bentz}},
  \bibinfo {author} {\bibfnamefont {I.~C.}\ \bibnamefont {Cloët}}\ and\
  \bibinfo {author} {\bibfnamefont {A.~W.}\ \bibnamefont {Thomas}},\ }\href
  {\doibase 10.1103/PhysRevD.85.014021} {\bibfield  {journal} {\bibinfo
  {journal} {Phys. Rev.}\ }\textbf {\bibinfo {volume} {D85}},\ \bibinfo {pages}
  {014021} (\bibinfo {year} {2012})}\Eprint {http://arxiv.org/abs/1111.1740}
  {~[arXiv:1111.1740 [hep-ph]]}\BibitemShut {NoStop}%
\bibitem [{\citenamefont {Cloët}\ \emph {et~al.}(2012)\citenamefont {Cloët},
  \citenamefont {Bentz},\ and\ \citenamefont {Thomas}}]{Cloet:2012td}%
  \BibitemOpen
  \bibfield  {author} {\bibinfo {author} {\bibfnamefont {I.~C.}\ \bibnamefont
  {Cloët}}, \bibinfo {author} {\bibfnamefont {W.}~\bibnamefont {Bentz}}\ and\
  \bibinfo {author} {\bibfnamefont {A.~W.}\ \bibnamefont {Thomas}},\ }\href
  {\doibase 10.1103/PhysRevLett.109.182301} {\bibfield  {journal} {\bibinfo
  {journal} {Phys. Rev. Lett.}\ }\textbf {\bibinfo {volume} {109}},\ \bibinfo
  {pages} {182301} (\bibinfo {year} {2012})}\Eprint
  {http://arxiv.org/abs/1202.6401} {~[arXiv:1202.6401 [nucl-th]]}\BibitemShut
  {NoStop}%
\bibitem [{\citenamefont {Hutauruk}\ \emph {et~al.}(2016)\citenamefont
  {Hutauruk}, \citenamefont {Cloët},\ and\ \citenamefont
  {Thomas}}]{Hutauruk:2016sug}%
  \BibitemOpen
  \bibfield  {author} {\bibinfo {author} {\bibfnamefont {P.~T.~P.}\
  \bibnamefont {Hutauruk}}, \bibinfo {author} {\bibfnamefont {I.~C.}\
  \bibnamefont {Cloët}}\ and\ \bibinfo {author} {\bibfnamefont {A.~W.}\
  \bibnamefont {Thomas}},\ }\href {\doibase 10.1103/PhysRevC.94.035201}
  {\bibfield  {journal} {\bibinfo  {journal} {Phys. Rev.}\ }\textbf {\bibinfo
  {volume} {C94}},\ \bibinfo {pages} {035201} (\bibinfo {year} {2016})}\Eprint
  {http://arxiv.org/abs/1604.02853} {~[arXiv:1604.02853 [nucl-th]]}\BibitemShut
  {NoStop}%
\bibitem [{\citenamefont {Ninomiya}\ \emph {et~al.}(2017)\citenamefont
  {Ninomiya}, \citenamefont {Bentz},\ and\ \citenamefont
  {Cloët}}]{Ninomiya:2017ggn}%
  \BibitemOpen
  \bibfield  {author} {\bibinfo {author} {\bibfnamefont {Y.}~\bibnamefont
  {Ninomiya}}, \bibinfo {author} {\bibfnamefont {W.}~\bibnamefont {Bentz}}\
  and\ \bibinfo {author} {\bibfnamefont {I.~C.}\ \bibnamefont {Cloët}},\
  }\href {\doibase 10.1103/PhysRevC.96.045206} {\bibfield  {journal} {\bibinfo
  {journal} {Phys. Rev.}\ }\textbf {\bibinfo {volume} {C96}},\ \bibinfo {pages}
  {045206} (\bibinfo {year} {2017})}\Eprint {http://arxiv.org/abs/1707.03787}
  {~[arXiv:1707.03787 [nucl-th]]}\BibitemShut {NoStop}%
\bibitem [{\citenamefont {Bernard}\ and\ \citenamefont
  {Meissner} (1989)}]{Bernard:1989fe}%
  \BibitemOpen
  \bibfield  {author} {\bibinfo {author} {\bibfnamefont {V.}~\bibnamefont
  {Bernard}}\ and\ \bibinfo {author} {\bibfnamefont {U.~G.}\ \bibnamefont {Meissner}},\
  }\href {\doibase 10.1103/PhysRevC.39.2054} {\bibfield  {journal} {\bibinfo
  {journal} {Phys. Rev.}\ }\textbf {\bibinfo {volume} {C39}},\ \bibinfo {pages}
  {2054} (\bibinfo {year} {1989})}\BibitemShut {NoStop}%
\bibitem [{\citenamefont {Bernard}\ and\ \citenamefont
  {Meissner} (1988)}]{Bernard:1988bx}%
  \BibitemOpen
  \bibfield  {author} {\bibinfo {author} {\bibfnamefont {V.}~\bibnamefont
  {Bernard}}\ and\ \bibinfo {author} {\bibfnamefont {U.~G.}\ \bibnamefont {Meissner}},\
  }\href {\doibase 10.1103/PhysRevLett.61.2973.3, 10.1103/PhysRevLett.61.2296} {\bibfield  {journal} {\bibinfo
  {journal} {Phys. Rev. Lett.}\ }\textbf {\bibinfo {volume} {61}},\ \bibinfo {pages}
  {2296} (\bibinfo {year} {1988})}\BibitemShut {NoStop}%
\bibitem [{\citenamefont {Lemmer} (1995)}]{Lemmer:1995eb}%
  \BibitemOpen
  \bibfield  {author} {\bibinfo {author} {\bibfnamefont {R.~H.}~\bibnamefont
  {Lemmer}},\
  }\href {\doibase 10.1016/0375-9474(95)00331-T} {\bibfield  {journal} {\bibinfo
  {journal} {Nucl. Phys.}\ }\textbf {\bibinfo {volume} {A593}},\ \bibinfo {pages}
  {315} (\bibinfo {year} {1995})}\BibitemShut {NoStop}%
\bibitem [{\citenamefont {Schulze} (1994)}]{Schulze:1994fy}%
  \BibitemOpen
  \bibfield  {author} {\bibinfo {author} {\bibfnamefont {H.~J.}~\bibnamefont
  {Schulze}},\
  }\href {\doibase 10.1088/0954-3899/20/4/002} {\bibfield  {journal} {\bibinfo
  {journal} {J. Phys.}\ }\textbf {\bibinfo {volume} {G20}},\ \bibinfo {pages}
  {531} (\bibinfo {year} {1994})}\BibitemShut {NoStop}%
 \bibitem [{\citenamefont {Weiss}\ \emph {et~al.}(1993)\citenamefont {Weiss},
  \citenamefont {Buck}, \citenamefont {Alkofer},\ and\ \citenamefont
  {Reinhardt}}]{Weiss:1993kv}%
  \BibitemOpen
  \bibfield  {author} {\bibinfo {author} {\bibfnamefont {C.}~\bibnamefont
  {Weiss}}, \bibinfo {author} {\bibfnamefont {A.}~\bibnamefont {Buck}},
  \bibinfo {author} {\bibfnamefont {R.}~\bibnamefont {Alkofer}}\ and\ \bibinfo
  {author} {\bibfnamefont {H.}~\bibnamefont {Reinhardt}},\ }\href {\doibase
  10.1016/0370-2693(93)90477-Y} {\bibfield  {journal} {\bibinfo  {journal}
  {Phys. Lett.}\ }\textbf {\bibinfo {volume} {B312}},\ \bibinfo {pages} {6}
  (\bibinfo {year} {1993})}\BibitemShut {NoStop}%
  \bibitem [{\citenamefont {Blin}\ \emph {et~al.}(1988)\citenamefont
  {Blin}, \citenamefont {Hiller},\ and\ \citenamefont
  {Schaden}}]{Blin:1987hw}%
  \BibitemOpen
  \bibfield  {author} {\bibinfo {author} {\bibfnamefont {A.~H.}~
  \bibnamefont {Blin}}, \bibinfo {author} {\bibfnamefont {B.~}\
  \bibnamefont {Hiller}}\ and\ \bibinfo {author} {\bibfnamefont {M.~}\
  \bibnamefont {Schaden}},\ }\href {\doibase }
  {\bibfield  {journal} {\bibinfo  {journal} {Z. Phys.}\ }\textbf {\bibinfo
  {volume} {A331}},\ \bibinfo {pages} {75} (\bibinfo {year} {1988})}\BibitemShut
  {NoStop}%
 \bibitem [{\citenamefont {Theussl}\ \emph {et~al.}(2004)\citenamefont
  {Theussl}, \citenamefont {Noguera},\ and\ \citenamefont
  {Vento}}]{Theussl:2002xp}%
  \BibitemOpen
  \bibfield  {author} {\bibinfo {author} {\bibfnamefont {L.~}\
  \bibnamefont {Theussl}}, \bibinfo {author} {\bibfnamefont {S.~}\
  \bibnamefont {Noguera}}\ and\ \bibinfo {author} {\bibfnamefont {V.~}\
  \bibnamefont {Vento}},\ }\href {\doibase 10.1140/epja/i2003-10174-3}
  {\bibfield  {journal} {\bibinfo  {journal} {Eur. Phys. J.}\ }\textbf {\bibinfo
  {volume} {A20}},\ \bibinfo {pages} {483} (\bibinfo {year} {2004})}\BibitemShut
  {NoStop}%
   \bibitem [{\citenamefont {Noguera}\ and\ \citenamefont
  {Scopetta} (2015)}]{Noguera:2015iia}%
  \BibitemOpen
  \bibfield  {author} {\bibinfo {author} {\bibfnamefont {S.}~\bibnamefont
  {Noguera}}\ and\ \bibinfo {author} {\bibfnamefont {S.}~ \bibnamefont {Scopetta}},\
  }\href {\doibase 10.1007/JHEP11(2015)102} {\bibfield  {journal} {\bibinfo
  {journal} {J. High Energy Phys.}\ }\textbf {\bibinfo {volume} {1511}},\ \bibinfo {pages}
  {102} (\bibinfo {year} {2015})}\BibitemShut {NoStop}%
 \bibitem [{\citenamefont {Courtoy}\ and\ \citenamefont
  {Naguera} (2007)}]{Courtoy:2007vy}%
  \BibitemOpen
  \bibfield  {author} {\bibinfo {author} {\bibfnamefont {A.}~\bibnamefont
  {Courtoy}}\ and\ \bibinfo {author} {\bibfnamefont {S.}~ \bibnamefont {Noguera}},\
  }\href {\doibase 10.1103/PhysRevD.76.094026} {\bibfield  {journal} {\bibinfo
  {journal} {Phys. Rev.}\ }\textbf {\bibinfo {volume} {D76}},\ \bibinfo {pages}
  {094026} (\bibinfo {year} {2007})}\BibitemShut {NoStop}%
  \bibitem [{\citenamefont {Hippe} (1995)}]{Hippe:1995hu}%
  \BibitemOpen
  \bibfield  {author} {\bibinfo {author} {\bibfnamefont {H.~J.}~\bibnamefont
  {Hippe}},\ and\ \bibinfo {author} {\bibfnamefont {S. P.}~ \bibnamefont {Klevansky}}
  }\href {\doibase 10.1103/PhysRevC.52.2172} {\bibfield  {journal} {\bibinfo
  {journal} {Phys. Rev.}\ }\textbf {\bibinfo {volume} {C52}},\ \bibinfo {pages}
  {2172} (\bibinfo {year} {1995})}\BibitemShut {NoStop}%
   \bibitem [{\citenamefont {Shigetani}\ \emph {et~al.}(1993)\citenamefont
  {Shigetani}, \citenamefont {Suzuki},\ and\ \citenamefont
  {Toki}}]{Shigetani:1993dx}%
  \BibitemOpen
  \bibfield  {author} {\bibinfo {author} {\bibfnamefont {T.~}\
  \bibnamefont {Shigetani}}, \bibinfo {author} {\bibfnamefont {K.~}\
  \bibnamefont {Suzuki}}\ and\ \bibinfo {author} {\bibfnamefont {H.~}\
  \bibnamefont {Toki}},\ }\href {\doibase 10.1016/0370-2693(93)91302-4}
  {\bibfield  {journal} {\bibinfo  {journal} {Phys. Lett.}\ }\textbf {\bibinfo
  {volume} {B308}},\ \bibinfo {pages} {383} (\bibinfo {year} {1993})}\BibitemShut
  {NoStop}%
   \bibitem [{\citenamefont {Davidson}\ and\ \citenamefont
  {Ruiz Arriola} (1995)}]{Davidson:1994uv}%
  \BibitemOpen
  \bibfield  {author} {\bibinfo {author} {\bibfnamefont {R.~M.}~\bibnamefont
  {Davidson}}\ and\ \bibinfo {author} {\bibfnamefont {E.~Ruiz}~ \bibnamefont {Arriola}},\
  }\href {\doibase 10.1016/0370-2693(95)00091-X} {\bibfield  {journal} {\bibinfo
  {journal} {Phys. Lett.}\ }\textbf {\bibinfo {volume} {B348}},\ \bibinfo {pages}
  {163} (\bibinfo {year} {1995})}\BibitemShut {NoStop}%
   \bibitem [{\citenamefont {Davidson}\ and\ \citenamefont
  {Ruiz Arriola} (2001)}]{Davidson:2001cc}%
  \BibitemOpen
  \bibfield  {author} {\bibinfo {author} {\bibfnamefont {R.~M.}~\bibnamefont
  {Davidson}}\ and\ \bibinfo {author} {\bibfnamefont {E.~Ruiz}~ \bibnamefont {Arriola}},\
  }\href {\doibase } {\bibfield  {journal} {\bibinfo
  {journal} {Acta. Phys. Polon.}\ }\textbf {\bibinfo {volume} {B33}},\ \bibinfo {pages}
  {1791} (\bibinfo {year} {2001})}\BibitemShut {NoStop}%
  \bibitem [{\citenamefont {Arriola} (2002)}]{RuizArriola:2002wr}%
  \BibitemOpen
  \bibfield  {author} {\bibinfo {author} {\bibfnamefont {E.~Ruiz.}~\bibnamefont
  {Arriola}},\
  }\href {\doibase } {\bibfield  {journal} {\bibinfo
  {journal} {Acta. Phys. Polon.}\ }\textbf {\bibinfo {volume} {B33}},\ \bibinfo {pages}
  {4443} (\bibinfo {year} {2002})}\BibitemShut {NoStop}%
 \bibitem [{\citenamefont {Dmitrasinovic}\ \emph {et~al.}(1992)\citenamefont
  {Dmitrasinovic}, \citenamefont {Lemmer},\ and\ \citenamefont
  {Tegen}}]{Dmitrasinovic:1992hb}%
  \BibitemOpen
  \bibfield  {author} {\bibinfo {author} {\bibfnamefont {V.~}\
  \bibnamefont {Dmitrasinovic}}, \bibinfo {author} {\bibfnamefont {R.~H.}~
  \bibnamefont {Lemmer}}\ and\ \bibinfo {author} {\bibfnamefont {R.~}\
  \bibnamefont {Tegen}},\ }\href {\doibase 10.1016/0370-2693(92)90420-9}
  {\bibfield  {journal} {\bibinfo  {journal} {Phys. Lett.}\ }\textbf {\bibinfo
  {volume} {B284}},\ \bibinfo {pages} {201} (\bibinfo {year} {1992})}\BibitemShut
  {NoStop}%
\bibitem [{\citenamefont {Beringer}\ \emph {et~al.}(2012)\citenamefont
  {Beringer} \emph {et~al.}}]{Beringer:1900zz}%
  \BibitemOpen
  \bibfield  {author} {\bibinfo {author} {\bibfnamefont {J.}~\bibnamefont
  {Beringer}} \emph {et~al.} (\bibinfo {collaboration} {Particle Data Group}),\
  }\href {\doibase 10.1103/PhysRevD.86.010001} {\bibfield  {journal} {\bibinfo
  {journal} {Phys. Rev. D}\ }\textbf {\bibinfo {volume} {86}},\ \bibinfo
  {pages} {010001} (\bibinfo {year} {2012})}\BibitemShut {NoStop}%
\bibitem [{\citenamefont {Durr}\ \emph {et~al.}(2011)\citenamefont {Durr},
  \citenamefont {Fodor}, \citenamefont {Hoelbling}, \citenamefont {Katz},
  \citenamefont {Krieg}, \citenamefont {Kurth}, \citenamefont {Lellouch},
  \citenamefont {Lippert}, \citenamefont {Szabo},\ and\ \citenamefont
  {Vulvert}}]{Durr:2010vn}%
  \BibitemOpen
  \bibfield  {author} {\bibinfo {author} {\bibfnamefont {S.}~\bibnamefont
  {Durr}}, \bibinfo {author} {\bibfnamefont {Z.}~\bibnamefont {Fodor}},
  \bibinfo {author} {\bibfnamefont {C.}~\bibnamefont {Hoelbling}}, \bibinfo
  {author} {\bibfnamefont {S.~D.}\ \bibnamefont {Katz}}, \bibinfo {author}
  {\bibfnamefont {S.}~\bibnamefont {Krieg}}, \bibinfo {author} {\bibfnamefont
  {T.}~\bibnamefont {Kurth}}, \bibinfo {author} {\bibfnamefont
  {L.}~\bibnamefont {Lellouch}}, \bibinfo {author} {\bibfnamefont
  {T.}~\bibnamefont {Lippert}}, \bibinfo {author} {\bibfnamefont {K.~K.}\
  \bibnamefont {Szabo}}\ and\ \bibinfo {author} {\bibfnamefont
  {G.}~\bibnamefont {Vulvert}},\ }\href {\doibase
  10.1016/j.physletb.2011.05.053} {\bibfield  {journal} {\bibinfo  {journal}
  {Phys. Lett.}\ }\textbf {\bibinfo {volume} {B701}},\ \bibinfo {pages} {265}
  (\bibinfo {year} {2011})}\Eprint {http://arxiv.org/abs/1011.2403}
  {~[arXiv:1011.2403 [hep-lat]]}\BibitemShut {NoStop}%
\bibitem [{\citenamefont {Patrignani}\ \emph {et~al.}(2016)\citenamefont
  {Patrignani} \emph {et~al.}}]{Patrignani:2016xqp}%
  \BibitemOpen
  \bibfield  {author} {\bibinfo {author} {\bibfnamefont {C.}~\bibnamefont
  {Patrignani}} \emph {et~al.} (\bibinfo {collaboration} {Particle Data
  Group}),\ }\href {\doibase 10.1088/1674-1137/40/10/100001} {\bibfield
  {journal} {\bibinfo  {journal} {Chin. Phys.}\ }\textbf {\bibinfo {volume}
  {C40}},\ \bibinfo {pages} {100001} (\bibinfo {year} {2016})}\BibitemShut
  {NoStop}%
\bibitem [{\citenamefont {Farrar}\ and\ \citenamefont
  {Jackson}(1979)}]{Farrar:1979aw}%
  \BibitemOpen
  \bibfield  {author} {\bibinfo {author} {\bibfnamefont {G.~R.}\ \bibnamefont
  {Farrar}}\ and\ \bibinfo {author} {\bibfnamefont {D.~R.}\ \bibnamefont
  {Jackson}},\ }\href {\doibase 10.1103/PhysRevLett.43.246} {\bibfield
  {journal} {\bibinfo  {journal} {Phys. Rev. Lett.}\ }\textbf {\bibinfo
  {volume} {43}},\ \bibinfo {pages} {246} (\bibinfo {year} {1979})}\BibitemShut
  {NoStop}%
\bibitem [{\citenamefont {Lepage}\ and\ \citenamefont
  {Brodsky}(1979)}]{Lepage:1979zb}%
  \BibitemOpen
  \bibfield  {author} {\bibinfo {author} {\bibfnamefont {G.~P.}\ \bibnamefont
  {Lepage}}\ and\ \bibinfo {author} {\bibfnamefont {S.~J.}\ \bibnamefont
  {Brodsky}},\ }\href {\doibase 10.1016/0370-2693(79)90554-9} {\bibfield
  {journal} {\bibinfo  {journal} {Phys. Lett. B}\ }\textbf {\bibinfo {volume}
  {87}},\ \bibinfo {pages} {359} (\bibinfo {year} {1979})}\BibitemShut
  {NoStop}%
\bibitem [{\citenamefont {Lepage}\ and\ \citenamefont
  {Brodsky}(1980)}]{Lepage:1980fj}%
  \BibitemOpen
  \bibfield  {author} {\bibinfo {author} {\bibfnamefont {G.~P.}\ \bibnamefont
  {Lepage}}\ and\ \bibinfo {author} {\bibfnamefont {S.~J.}\ \bibnamefont
  {Brodsky}},\ }\href {\doibase 10.1103/PhysRevD.22.2157} {\bibfield  {journal}
  {\bibinfo  {journal} {Phys. Rev.}\ }\textbf {\bibinfo {volume} {D22}},\
  \bibinfo {pages} {2157} (\bibinfo {year} {1980})}\BibitemShut {NoStop}%
\bibitem [{\citenamefont {Chang}\ \emph {et~al.}(2013)\citenamefont {Chang},
  \citenamefont {Cloët}, \citenamefont {Roberts}, \citenamefont {Schmidt},\
  and\ \citenamefont {Tandy}}]{Chang:2013nia}%
  \BibitemOpen
  \bibfield  {author} {\bibinfo {author} {\bibfnamefont {L.}~\bibnamefont
  {Chang}}, \bibinfo {author} {\bibfnamefont {I.~C.}\ \bibnamefont {Cloët}},
  \bibinfo {author} {\bibfnamefont {C.~D.}\ \bibnamefont {Roberts}}, \bibinfo
  {author} {\bibfnamefont {S.~M.}\ \bibnamefont {Schmidt}}\ and\ \bibinfo
  {author} {\bibfnamefont {P.~C.}\ \bibnamefont {Tandy}},\ }\href {\doibase
  10.1103/PhysRevLett.111.141802} {\bibfield  {journal} {\bibinfo  {journal}
  {Phys. Rev. Lett.}\ }\textbf {\bibinfo {volume} {111}},\ \bibinfo {pages}
  {141802} (\bibinfo {year} {2013})}\Eprint {http://arxiv.org/abs/1307.0026}
  {~[arXiv:1307.0026 [nucl-th]]}\BibitemShut {NoStop}%
\bibitem [{\citenamefont {Cloët}\ \emph {et~al.}(2013)\citenamefont {Cloët},
  \citenamefont {Chang}, \citenamefont {Roberts}, \citenamefont {Schmidt},\
  and\ \citenamefont {Tandy}}]{Cloet:2013tta}%
  \BibitemOpen
  \bibfield  {author} {\bibinfo {author} {\bibfnamefont {I.~C.}\ \bibnamefont
  {Cloët}}, \bibinfo {author} {\bibfnamefont {L.}~\bibnamefont {Chang}},
  \bibinfo {author} {\bibfnamefont {C.~D.}\ \bibnamefont {Roberts}}, \bibinfo
  {author} {\bibfnamefont {S.~M.}\ \bibnamefont {Schmidt}}\ and\ \bibinfo
  {author} {\bibfnamefont {P.~C.}\ \bibnamefont {Tandy}},\ }\href {\doibase
  10.1103/PhysRevLett.111.092001} {\bibfield  {journal} {\bibinfo  {journal}
  {Phys. Rev. Lett.}\ }\textbf {\bibinfo {volume} {111}},\ \bibinfo {pages}
  {092001} (\bibinfo {year} {2013})}\Eprint {http://arxiv.org/abs/1306.2645}
  {~[arXiv:1306.2645 [nucl-th]]}\BibitemShut {NoStop}%
\bibitem [{\citenamefont {Barone}\ \emph {et~al.}(2002)\citenamefont {Barone},
  \citenamefont {Drago},\ and\ \citenamefont {Ratcliffe}}]{Barone:2001sp}%
  \BibitemOpen
  \bibfield  {author} {\bibinfo {author} {\bibfnamefont {V.}~\bibnamefont
  {Barone}}, \bibinfo {author} {\bibfnamefont {A.}~\bibnamefont {Drago}}\ and\
  \bibinfo {author} {\bibfnamefont {P.~G.}\ \bibnamefont {Ratcliffe}},\ }\href
  {\doibase 10.1016/S0370-1573(01)00051-5} {\bibfield  {journal} {\bibinfo
  {journal} {Phys. Rept.}\ }\textbf {\bibinfo {volume} {359}},\ \bibinfo
  {pages} {1} (\bibinfo {year} {2002})}\Eprint
  {http://arxiv.org/abs/hep-ph/0104283} {~[arXiv:hep-ph/0104283
  [hep-ph]]}\BibitemShut {NoStop}%
\bibitem [{\citenamefont {Bertone}\ \emph {et~al.}(2014)\citenamefont
  {Bertone}, \citenamefont {Carrazza},\ and\ \citenamefont
  {Rojo}}]{Bertone:2013vaa}%
  \BibitemOpen
  \bibfield  {author} {\bibinfo {author} {\bibfnamefont {V.}~\bibnamefont
  {Bertone}}, \bibinfo {author} {\bibfnamefont {S.}~\bibnamefont {Carrazza}}\
  and\ \bibinfo {author} {\bibfnamefont {J.}~\bibnamefont {Rojo}},\ }\href
  {\doibase 10.1016/j.cpc.2014.03.007} {\bibfield  {journal} {\bibinfo
  {journal} {Comput. Phys. Commun.}\ }\textbf {\bibinfo {volume} {185}},\
  \bibinfo {pages} {1647} (\bibinfo {year} {2014})}\Eprint
  {http://arxiv.org/abs/1310.1394} {~[arXiv:1310.1394 [hep-ph]]}\BibitemShut
  {NoStop}%
\bibitem [{\citenamefont {Cloët}\ \emph {et~al.}(2005)\citenamefont {Cloët},
  \citenamefont {Bentz},\ and\ \citenamefont {Thomas}}]{Cloet:2005pp}%
  \BibitemOpen
  \bibfield  {author} {\bibinfo {author} {\bibfnamefont {I.~C.}\ \bibnamefont
  {Cloët}}, \bibinfo {author} {\bibfnamefont {W.}~\bibnamefont {Bentz}}\ and\
  \bibinfo {author} {\bibfnamefont {A.~W.}\ \bibnamefont {Thomas}},\ }\href
  {\doibase 10.1016/j.physletb.2005.06.065} {\bibfield  {journal} {\bibinfo
  {journal} {Phys. Lett.}\ }\textbf {\bibinfo {volume} {B621}},\ \bibinfo
  {pages} {246} (\bibinfo {year} {2005})}\Eprint
  {http://arxiv.org/abs/hep-ph/0504229} {~[arXiv:hep-ph/0504229
  [hep-ph]]}\BibitemShut {NoStop}%
\end{thebibliography}

%

\end{document}